\documentclass[fleqn,usenatbib]{mnras}

\usepackage{newtxtext,newtxmath}

\usepackage[T1]{fontenc}
\usepackage{MnSymbol}
\DeclareRobustCommand{\VAN}[3]{#2}
\let\VANthebibliography\thebibliography
\def\thebibliography{\DeclareRobustCommand{\VAN}[3]{##3}\VANthebibliography}

\usepackage{graphicx}	
\usepackage{amsmath}	
\usepackage{comment}

\title[Low Spin of Black Holes Embedded in Turbulent AGN Discs]{Chaotic Gas Accretion by Black Holes Embedded in AGN Discs as Cause of Low-spin Signatures in Gravitational Wave Events}

\author[Chen \& Lin]{Yi-Xian Chen $^{1}$\thanks{yc9993@princeton.edu}
and 
Douglas N. C. Lin$^{2,3}$
\\
$^{1}${Department of Astrophysical Sciences, Princeton University, USA}
\\
$^{2}${Department of Astronomy \& Astrophysics, University of California, Santa Cruz, CA 95064, USA}
\\
$^{3}${Institute for Advanced Studies, Tsinghua University, Beijing 100084, China}
}

\date{Accepted XXX. Received YYY; in original form ZZZ}

\pubyear{2015}

\begin{document}
\label{firstpage}
\pagerange{\pageref{firstpage}--\pageref{lastpage}}
\maketitle

\begin{abstract}
Accretion discs around super-massive black holes (SMBH) not only power active galactic nuclei (AGNs),
but also host single and binary embedded stellar-mass black holes (EBHs) that grow rapidly from gas accretion. The merger of these EBHs provides
a promising mechanism for the excitation of some gravitational wave events observed by LIGO-Virgo, especially those with source masses considerably larger than isolated stellar-mass black hole binaries. 
In addition to their mass and 
mass-ratio distribution, {their hitherto enigmatic small spin-parameters 
($\chi_{\rm eff}$)} 
carry important clues and stringent constraints on their formation channels and evolutionary pathways. Here we show that, {between each coalescence, the typical rapid spin of the merged EBHs
is suppressed by their subsequent accretion of gas from a turbulent environment, 
due to its ability to randomize the flow's spin orientation with respect to 
that of the EBHs on an eddy-turnover timescale.}
This theory provides supporting evidence for the prolificacy of EBH mergers
and suggests that their mass growth may be dominated by gas accretion rather than their coalescence 
in AGN discs. 
\end{abstract}




\section{Introduction}

Direct observation of the center of our milky way \citep{ghez2003, bartko2010}, 
as well as abundant tidal disruption event samples 
\citep{Law-Smith2017, mockler2021} suggest that stellar clusters commonly exist around super-massive black holes (SMBHs) \citep{KormendyHo2013}. 
In active galactic nuclei (AGNs),
the cluster stars may be captured into circularized orbits on the SMBH accretion disc midplane through resonance coupling and gas drag during disc 
passage \citep{Artymowicz1993, Macloed2020}. 
Embedded stars may also 
form \textit{in situ} from gravitational instability \citep{Goodman2003,Jiang2011,Stone2017,Chen2023}.
Due to the rapid accretion of disc material, they evolve quickly to become massive stars \citep{Cantiello2021} and 
then undergo supernova or gravitational collapse, leaving behind not only ejecta that might account for metallicity abundances in AGNs \citep{Hamann1999,Hamann2002}, 
but also embedded stellar mass black holes (EBHs). {Binary stellar-mass black holes (BBHs) may form
through dynamical encounters during the global ramp-up of stand-alone EBHs 
or their local accumulation
at migration traps \citep{Bellovary2016}. 
The hardening and eventual merging of these binaries can be a promising channel 
to produce gravitational waves (GW) that contribute 
to LIGO-Virgo events  \citep{McKernan2012, McKernan2014,Yang2019,Tagawa2020,Samsing2022}, especially those with progenitor masses being considerably larger than isolated stellar-mass black hole binaries.} Additionally,
BBH mergers in an 
AGN disc may shock-heat the surrounding accretion flow and generate optical/UV flares, an electromagnetic counterpart that could 
differentiate them from other merger channels \citep{graham2020,Veronesi2022}. 

The coalescence of two comparable-mass EBHs generally leads to a merged product with a combined 
mass $M_\filledstar$ and large spin angular momentum $J_\filledstar$ \citep{dones1993, hofmann2016}. 
In a laminar global AGN disc, EBHs with circularized orbits around the SMBH can gain spin 
angular momentum between merger events through gas accretion from 
their local circum-stellar discs (CSDs), and quickly become EBHs with high spin aligned with each other (prograde to the disc 
rotation), even if their initial spins can be negligible \citep{FullerMa2019}.  
However, the 
projection of the mass-weighted spin-angular-momentum of individual EBHs in the BBHs' orbital 
angular momentum direction ($\chi_{\rm eff}$),
inferred from the observed GW events, prefers low values \citep{LIGO2021-third2}.  
This distribution 
suggests low natal EBH spins or random directions between the binary orbital angular momentum and 
the EBHs' individual spins \citep{Farr2017}.

While dynamical encounters between BBHs/EBHs and
other stars can tilt their orbital planes significantly away from the disc plane, 
this effect alone cannot reduce the dispersion in $\chi_{\rm eff}$ down to typical observational values $<0.1$, 
and {frequent dynamical interactions 
tend to result in an EBH mass distribution skewed 
towards higher mass compared with observation \citep{Tagawa2020}, but see \citet{Tagawa2021} for mitigating this by disruption of soft binaries during binary–binary interactions}. 
It has also been suggested that EBHs or BBHs with non-negligible eccentricity are 
surrounded by CSDs
with retrograde spins \citep{liYP2022,chen2022prograde} which introduces misalignment between spin axes of merging EBHs, 
as possible solutions to this paradox. 
En route potential paths of BBHs' migration, evection 
and eviction resonances 
between their precession and orbital frequencies can excite eccentricity 
in BBHs with nearly co-planar and highly inclined orbits whereas 
spin-orbit resonances can also 
modify BBHs' obliquity \citep{2022arXiv220407282G}. 

{While fore-mentioned mechanisms rely on misalignment/counter-alignment between EBH populations to produce low-$\chi_{\rm eff}$ events, 
they pose no constraint on the growth of individual EBH spins.} {Simulations have shown that Blandford-Znajek jet can significantly contribute to spin down of isolated Black Holes \citep{Narayan2022}, although its effect on EBH spin distribution has not been explored in details. } On the other hand, sonic-scale magneto-rotational and gravitational instabilities (MRI and GI), commonly occur in AGN accretion discs \citep{BalbusHawley1998,Gammie2001,Goodman2003}. Both instabilities excite turbulence with locally
chaotic eddies. 
In this paper, we show that that EBHs' accretion from strongly turbulent eddies provides 
an alternative and novel mechanism to robustly reduce the dimensionless spin parameter $a = cJ_{\filledstar}/GM_{\filledstar}^2$
of individual EBHs before they capture {or after they merge with} their binary companions. 

This paper is organized as follows: In \S \ref{sec:simulation}, we present exemplary numerical simulations that provide insights into how turbulence can affect the spin of CSDs. In \S \ref{sec:individual_spin}, we introduce our prescribed models for the spin evolution of individual black holes. Then, in \S \ref{sec:Lense-Thirring}, we consider the additional effects of Lense-Thirring torques on spin reorientation and demonstrate that it is not significant in our parameter space of interest. We then apply these methods to study the spin distribution evolution of a population of EBHs in \S \ref{sec:pop} and discuss the implications of our findings in \S \ref{sec:conclusion}.

\section{Simulation of CSD Flow Disrupted by Turbulence}
\label{sec:simulation}
\subsection{Numerical Setup}

\begin{figure*}
\centering
\includegraphics[width=0.9\textwidth]{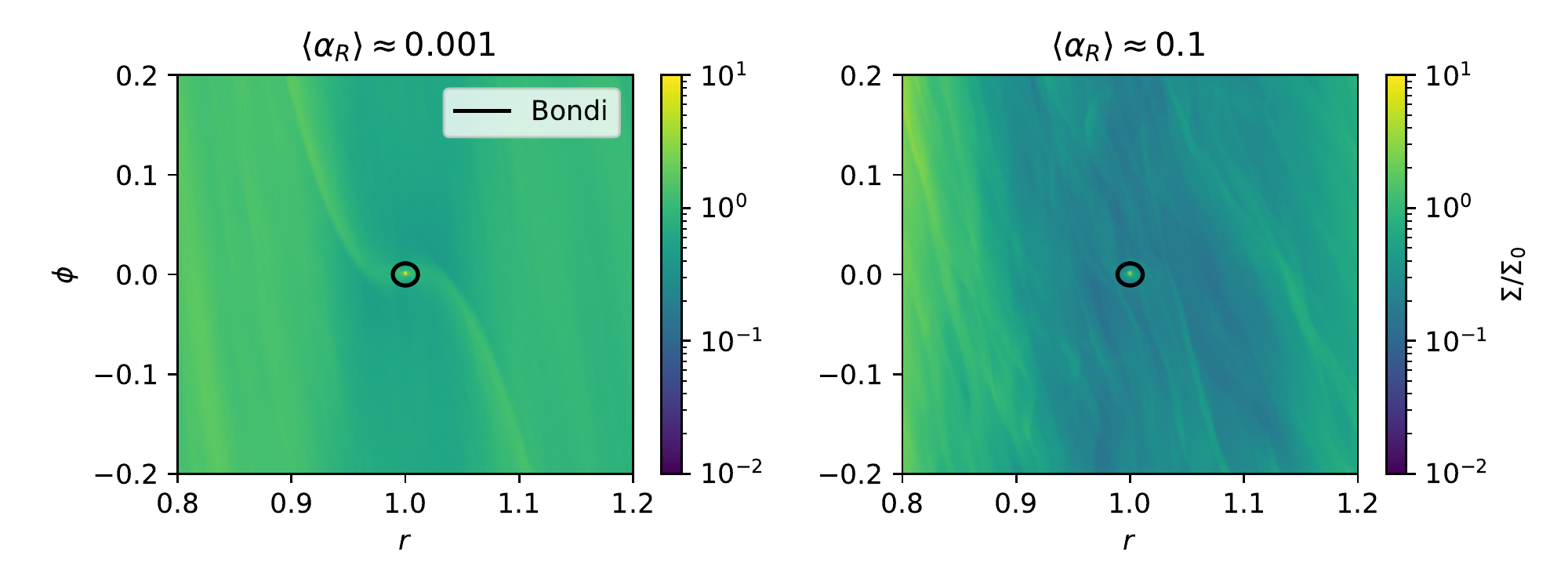}
\caption{Surface density distribution around the companion when simulation reaches quasi-steady state, averaged over 20 snapshots during one orbital timescale. The normalization density $\Sigma_0$ is the initial disc surface density at the companion location. The companion Bondi radius is shown in black lines.}
\label{fig:num_density}
\end{figure*}

\begin{figure*}
\centering
\includegraphics[width=1.0\textwidth]{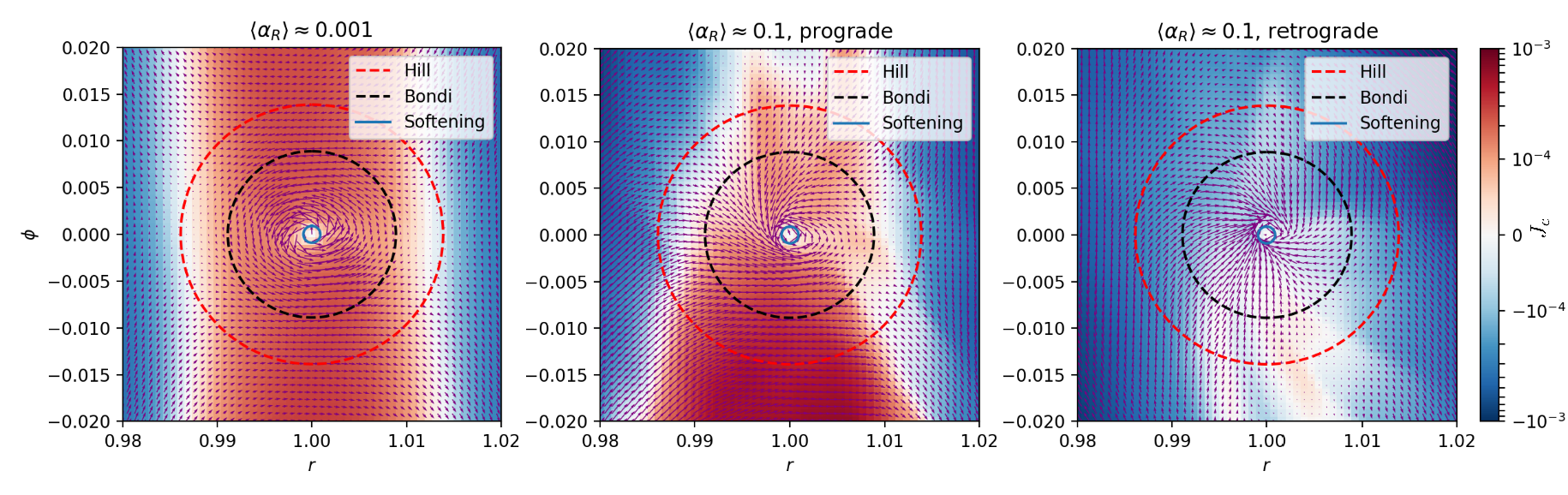}
\caption{Zoom-in distribution of specific angular momentum with respect to the companion $J_c$, also averaged over one orbital timescale. Effectively, red means prograde and blue means retrograde. Middle panel and right panel show flow patterns during characteristic prograde and retrograde cycles of CSD flow in the highly turbulent run. Small blue circle, black circle and red circle centered on the companion shows the softening length, Bondi and Hill radius, respectively.}
\label{fig:AM_plots}
\end{figure*}

{To explore the effect of turbulence on the spin re-orientation of embedded CSDs and determine characteristic values of  $\mathcal{S}$, 
we apply a modified version of the grid-based code FARGO \citep{Masset2000} 
with a phenomenological turbulence prescription that follows 
\citet{Laughlin2004} and  \citet{BaruteauLin}.}

For our initial conditions, 
we choose an axisymmetric 2D Keplerian disc model, 
with the aspect ratio only a function of distance $r$ to the SMBH :

\begin{equation}
    h=\frac{c_{s}}{\Omega r}=h_{0}\left(\frac{r}{r_{0}}\right)^{1 / 4}
\end{equation}

where $c_s$ is the sound speed, $\Omega$ is the Keplerian frequency and $r_0$ is the EBH’s orbital radius. 
The code unit is $G = M_{\bullet} = r_0 = 1$, where $M_{\bullet}$ is the central SMBH mass. The EBH's mass ratio is $M_* = qM_{\bullet}$, where $q=8\times 10^{-6}$ is our fiducial mass ratio. 
The EBH companion is placed on circular orbit at $(r,\phi) = (r_0, 0)$. 
We numerically solve the vertically-integrated hydrodynamic equations in a rotating frame centered on the host star with angular velocity $\Omega(r_0)$, 
stationary relative to the companion EBH.  The gas surface density profile is initialized as 

\begin{equation}
\Sigma=\Sigma_{ 0}\left(\frac{r}{r_{0}}\right)^{-1},
\end{equation}

We set $h_0 = 0.03$ and $\Sigma_0 = 1.0$, 
the latter is simply a normalization constant since we do not calculate gas feedback onto the EBH. 
Since we focus on the CSDs of EBH, 
the global radial gradients in $h$ and $\Sigma$ do not affect our results. 
In the companion vicinity $R_B \sim R_H \lesssim H$, where $H = h_0r_0$ is the global disc's scale height at $r_0$, 
$R_B = GM_*/c_s^2 = 0.009r_0$ is the Bondi radius, and $R_H = r_0(q/3)^{1/3} = 0.014r_0$ is the Hill radius. 
On a local scale, the resulting flow pattern could correspond to $q \approx 0.3 h_0^3$ scaled towards any $h_0$ \citep{Ormel2013}. 
This setup represents generally the category of sub-thermal ($R_H<H$) companions. 
For large SMBH mass $M_\bullet \sim 10^8 M_\odot$ and $h_0\gtrsim 0.01$ \citep{Sirko2003}, sub-thermal companions should represent the most common kind of EBHs, 
if $M_* \lesssim 100 M_\odot$ consistent with LIGO detections. 


To achieve high resolution for studying local physical processes as well as to capture global turbulence qualities, 
we apply a wedge-like computational domain that extends from $0.6r_0$ to $1.4r_0$ in the radial direction, 
and $-\pi/8$ to $\pi/8$ in
the azimuthal direction, 
both resolved by 1024 grids with linear spacing. 
The radial boundary conditions are fixed towards their initial values while the azimuthal boundary condition is periodic. 

To prevent gas velocity from diverging infinitely close to the EBH, 
we consider a smoothed EBH direct potential of the Plummer form, 
with a grid-scale softening length $\epsilon = 0.1R_B \approx 0.064 R_H$ enough to resolve rotation within the CSD but still much larger than the true 
innermost stable circular orbit (ISCO) of the EBH ${R}_{\rm isco}$. Within $\epsilon$ 
(at grid scale), 
the gravity is softened and gas density inevitably piles up. 
Realistically, the gas would be compressed onto scales $\sim {R}_{\rm isco} \ll \epsilon $. 
To mitigate
the artificial concentration effect on grid scale while leaving gas dynamics at larger scales unchanged by its presence, 
{we introduce a sink term to reduce the surface density gas within $\epsilon$ by a factor of $f$ over each numerical (Courant–Friedrichs–Lewy, CFL) timestep $\tau_{\rm CFL}$. 
The ratio $A\equiv f/\tau_{\rm CFL}$ factor is controlled to be a constant such that in steady state,
a mass removal rate $\int_{|\mathbf{R}|<\epsilon} A \Sigma dS$ is reached where $\mathbf{R}$ is the gas fluid's displacement vector from the companion and $dS$ is unit surface.} 
We set $A =16 \Omega$, 
but for a sink boundary $<0.1R_H$ the removal rate has been proven to converge for different $A$ and is usually used as a planetary accretion rate in planet-disc simulations \citep{TanigawaWatanabe2002,DAngelo2003}. However, since these rates are close to the Bondi rate \citep{Li2021gasgiant} which is much higher than the Eddington rate of EBHs \citep{liYP2022,Tagawa2022}, 
we expect strong radiation feedback/jets \citep{Jiang2014,Jiang2019} occurring at scales $\lesssim \epsilon$ but $\gg R_{\rm isco}$ to significantly recycle out most of that mass flux and leave behind an accretion rate comparable to a few tens to hundreds of Eddington rate that eventually reaches $R_{\rm isco}$. 
In our simulation we neglect these effects and assume they do not affect large scale flow structures. 

A turbulent potential $\Phi_{\rm turb} \propto \gamma$ is applied to the disc, 
corresponding to the superposition of 50 wave-like modes \citep{Laughlin2004} such that

\begin{equation}
    \Phi_{\text {turb }}(r, \varphi, t)=\gamma r^{2} \Omega^{2} \sum_{k=1}^{50} \Lambda_{k}(m_k; r, \varphi, t).
\end{equation}

where $\gamma$ is the dimensionless characteristic amplitude of turbulence. 
Each stochastic factor for the $k$-th mode $\Lambda_{k}$ has a wavenumber $m_k$ randomly drawn from a logarithmically uniform distribution between \footnote{since smaller wavenumbers cannot be accommodated by our wedge-like simulation domain} $m=8$ and the largest wavenumber $m_{\rm max}$ corresponding to wavelength $\sim H$ \footnote{This approximates the power spectrum of Graveto-turbulence with a decay below $H$ scale \citep{BoothClarke2019}. For MRI one can include up to grid scale turbulence depending on whether small-scale structures are of interest \citep{Laughlin2004}. We confirm that two $m_{\rm max}$ make little difference since large scale eddies has 
more influence on the CSD flow.}. This turbulence driver produces the power spectrum of a typical Kolmogorov cascade $m^{-5/3}$ up to $m_{\rm max}$, 
and an effective Reynold stress parameter $\langle \alpha_R \rangle$ around the companion location can be measured from the velocity fluctuations that relates to the turbulence amplitude as $\langle \alpha_R \rangle \approx 35 (\gamma/h_0)^2$. 
(see \citet{BaruteauLin} for details).

\subsection{Results and Implication for EBH Spin-reorientation}

We present results from simulations with different $\gamma$ values, corresponding to $\langle \alpha_R \rangle \approx 0.001 $ and $\langle \alpha_R \rangle \approx 0.1$ respectively. From Figure \ref{fig:num_density}, we see that $\Sigma$ fluctuation in the high turbulence model is much stronger than the low turbulence model. The S-shape density wave, or ``wake" induced by companion-disc interaction in a laminar disc 
\citep{OgilvieLubow2002} is apparent in the left panel but inconspicuous in the right panel. 

We present zoom-in plots of flow patterns in Figure \ref{fig:AM_plots}. 
Distribution of specific angular momentum $J_c =|\mathbf{R} \times \mathbf{v}_c| $ is plotted in the background of $\mathbf{v}_c$ vectors, representing gas velocity with respect to the companion. 
In the weak turbulence model (left panel), the steady orbit-averaged flow patterns are very similar to low viscosity or inviscid calculations of planet-disc interaction \citep{DAngelo2003, Tanigawa2012,Ormel2013}. 
At radii far from the companion the effectively retrograde (blue) Keplerian shear dominates, while closer to the companion weak velocity fluctuation cannot prevent planet tidal potential from establishing a 
prograde (red) CSD within $R_B$ that connects with horseshoe streamlines at larger azimuth. This means that it's reasonable to approximate low turbulence with a kinematic viscosity term in laminar fluid equations, 
as in most planet-disc simulations.

However, in the strong turbulence model corresponding to very large $\alpha_R \gtrsim 0.1$ that gravitational instability in AGN discs may provide \citep{Gammie2001,Goodman2003,deng2020}, this steady-state flow pattern is disrupted and the disc becomes far from laminar. 
On average during certain individual orbital timescales, 
the EBH gravity is still able to generate a rotating CSD within the Bondi radius, 
albeit the typical rotation direction can fluctuate from prograde to retrograde intermittently under influence of turbulence. The middle and right panels of \ref{fig:AM_plots}
show orbit-averaged result of flow pattern during a typical prograde CSD episode and a typical retrograde CSD episode. 

\begin{figure*}
\centering
\includegraphics[width=1.0\textwidth]{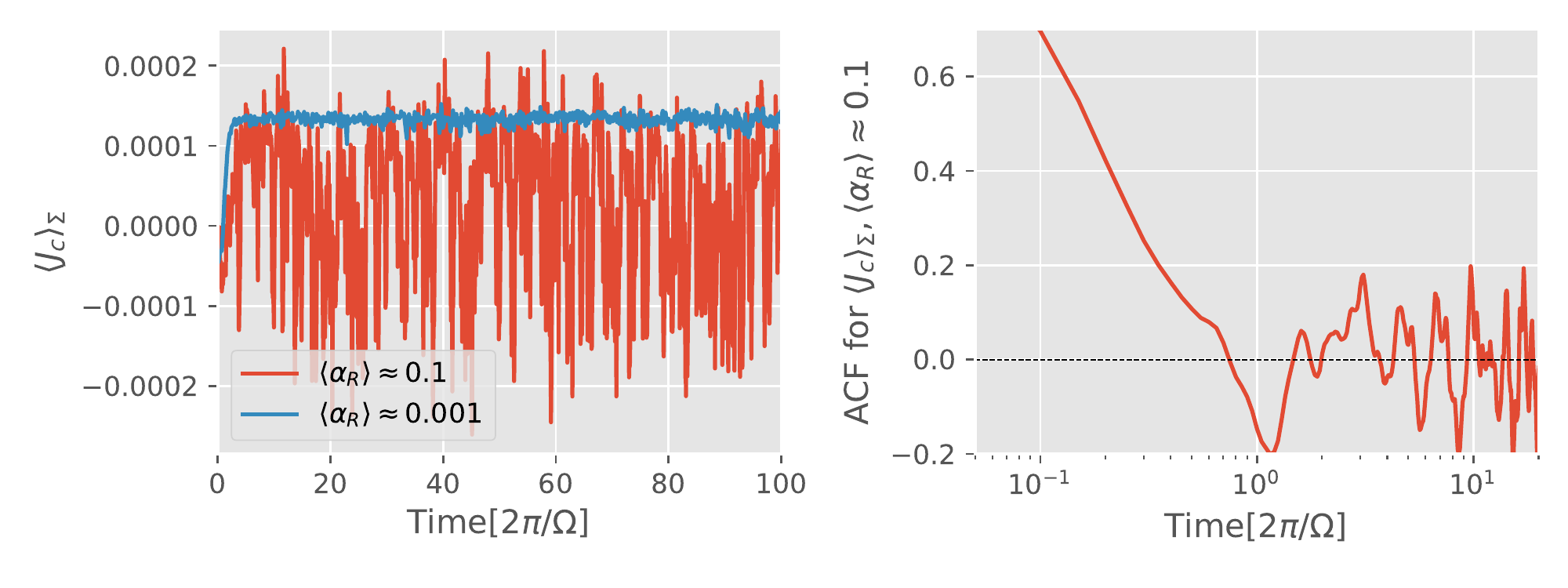}
\caption{Left: the mass-average specific angular momentum $\langle J_c\rangle_{\Sigma}$ within the Bondi radius as a function of time. Right: The auto-correlation function of the highly fluctuating $\langle J_c\rangle_{\Sigma}$ in the highly turbulent simulation, which shows the autocorrelation timescale is close to $2\pi/\Omega$.}
\label{fig:ave_J}
\end{figure*}

As an indicator for the general rotation direction of CSD, 
we calculate the mass-averaged specific angular momentum of gas within the companion Bondi radius 
(which contains materials instantaneously bound to the companion) excluding the small softened grid-scale region.

\begin{equation}
    \langle J_c\rangle_{\Sigma} = \dfrac{\int_{\epsilon<|\mathbf{R}| < R_B} J_c \Sigma dS}{\int_{\epsilon< |\mathbf{R}| <  R_B} \Sigma dS}
\end{equation}

and plot its time evolution in the left panel of Figure \ref{fig:ave_J}. 
In the weak turbulence model, $\langle J_c\rangle_{\Sigma}$ quickly settles towards a quasi-steady value on the order of $ \sim R_B c_s = 0.00027$ in code units, 
reflecting a steady prograde CSD within the Bondi radius.
In the strongly turbulent model, $\langle J_c\rangle_{\Sigma}$ fluctuates between positive and negative values, giving an overall time average an order-of-magnitude below $R_B c_s$. The right panel of Figure \ref{fig:ave_J} shows the normalized 
autocorrelation function (ACF) of the time-series $\langle J_c\rangle_{\Sigma}$, 

\begin{equation}
    \mathrm{ACF}(\tau) := \dfrac{\int_{t_{\rm min}+\tau}^{t_{\rm max}}\langle J_c\rangle_{\Sigma} (t-\tau) \langle J_c\rangle_{\Sigma}(t) dt}{\int_{t_{\rm min}+\tau}^{t_{\rm max}}\langle J_c\rangle_{\Sigma}^2(t)dt}
\end{equation}

We use $\langle J_c\rangle_{\Sigma}$ between $t_{\min} = 50$ and $t_{\max} = 100$ orbital timescales to calculate ACF. 
Measured from the ``second-zero-crossing" of ACF function \citep{Oishi2007,BaruteauLin}, the ACF timescale is equal to
$\Delta t_\filledstar
\sim 1-2 (2\pi/\Omega)$. 
Statistically, this indicates the typical timescale that rotation of CSD switches direction or the duration of each accretion episode with net angular momentum. 

{While the CSD flow shows a preferred rotation direction on average over a typical episode (as shown in Figure \ref{fig:AM_plots}, middle and right panels), the frequency of prograde and retrograde motion cancels out each other. Assuming that either prograde or retrograde rotation can be established down to $R_{\rm isco}$ during each episode, 
which is beyond the domain of our simulation, 
it's reasonable to speculate that the direction of specific angular momentum there, $J_c(R_{\rm isco})$, 
will also intermittently flip around due to large-scale flow changes, 
despite its magnitude remaining around the Keplerian value within each episode 
(see Eqn \ref{eqn:evolvea}, which constrain $|J_c({R}_{\rm isco})| \sim \sqrt{GM_{\filledstar}{R}_{\rm isco}}$). 
To confirm that the episodes cancel out, we define $f_J$ as a proxy to reflect the fractional imbalance between prograde and retrograde flow:}

\begin{equation}
    f_J = \dfrac{\int_{\langle J_c\rangle>0} dt - \int_{\langle J_c\rangle<0} dt}{\int dt }
    \label{eqn:imbalance}
\end{equation}

{which is the difference in the duration of prograde episodes v.s. retrograde episodes, 
normalized by the total duration. 
While for the laminar case it naturally converges to 100\%, 
we measure $f_J = 0.6\%$ in the turbulent case. The interpretation is that the rotation at small scales will be prograde $\approx 50.3\%$ of the time, 
while retrograde $\approx 49.7\%$ of the time. 
If we extrapolate that fraction to the flow rotation down to $R_{\rm isco}$, it
implies a non-stochastic component that would contribute to a net increase of $ \approx 0.6\%
$ in the value of $a$ in an Eddington timescale. 
As we will show in the next section, 
this contribution is smaller than the characteristic dispersion $\sim \mathcal{S}^{-0.5}$ contributed by Random Walk (RW). 
Based on the relatively small imbalance, 
we infer that 
in the strong-turbulence parameter space of interest, 
RW would dominate spin evolution in the long term.}

\begin{figure}
\centering
\includegraphics[width=0.45\textwidth]{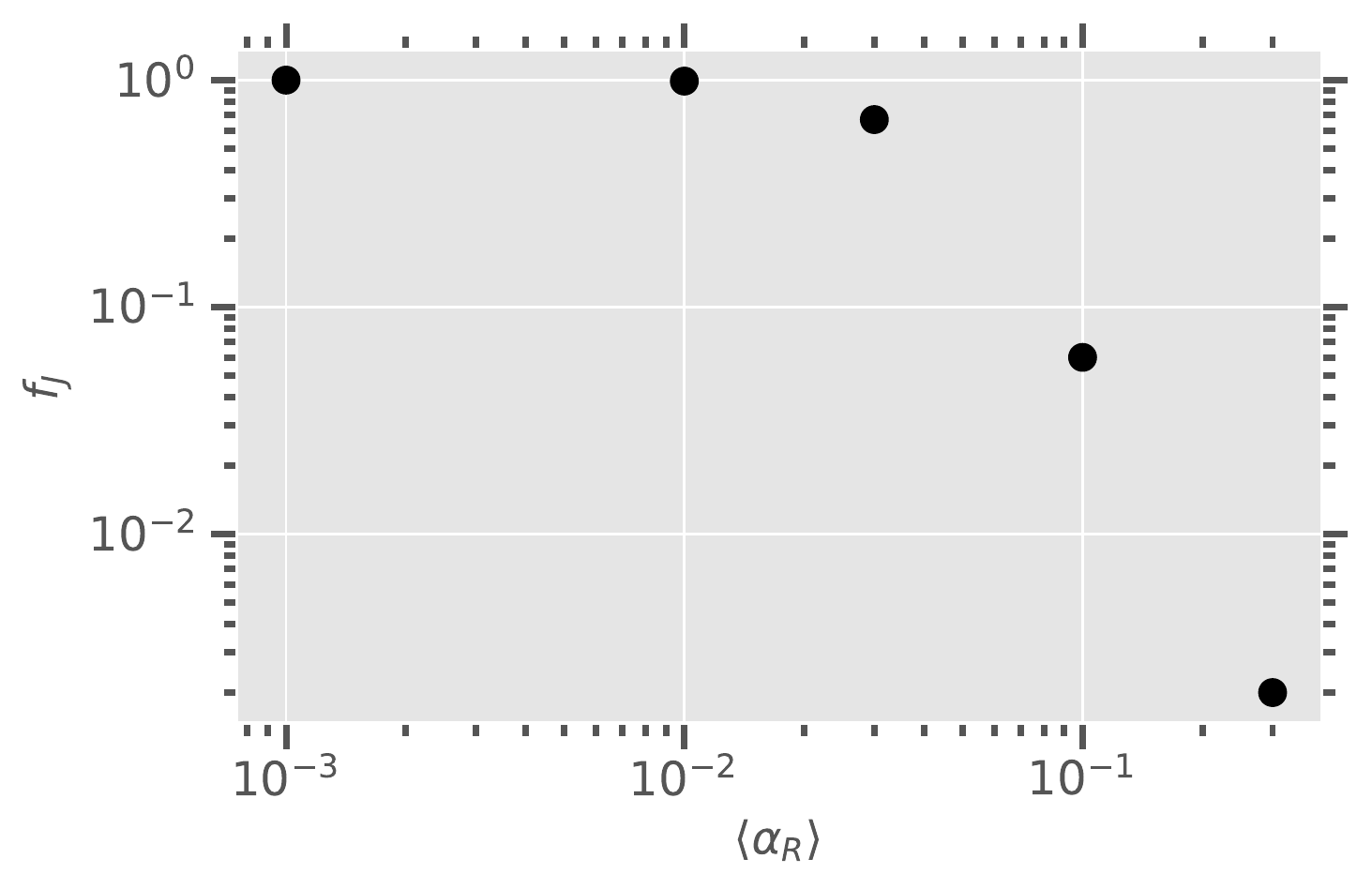}
\caption{The imbalance (Eqn \ref{eqn:imbalance}) between total duration of prograde episodes and retrograde episodes, as a function of $\langle \alpha_R \rangle$. 
This shows for our fiducial set of embedded-companion and disc parameters, 
the transition from orderly to stochastic CSD spin happens at $\langle \alpha_R \rangle \gtrsim 0.03$}
\label{fig:f_Jplot}
\end{figure}

{We plot $f_J$ from results of experiments with other values of $\langle \alpha_R \rangle$ in Figure \ref{fig:f_Jplot}, and conclude that for our fiducial set of parameters, $\langle\alpha_R \rangle \gtrsim 0.03$ appears to mark a phase transition from orderly to more stochastic accretion, and at $\alpha_R =0.3$ the re-orientation becomes highly random.
This transition may be related to fractional diffusion process 
(a fluctuating factor that gradually decays and a steady-state factor that gradually dominates), 
the details of which should be explored over larger parameter space by subsequent simulations, 
possibly with more realistic treatment of turbulence (e.g. with self-gravity and/or radiative cooling). }

We also explored another set of simulations with the same mass ratio but $h_0=0.01$, which has $R_B > R_H > H$ to represent highly super-thermal companions, common in thin discs $h_0\sim 0.001$ around low-mass SMBHs \citep{Tagawa2020a} $M_\bullet \sim 10^6 M_\odot$  or for intermediate mass EBHs \citep{McKernan2012}. 
In such cases steady-state prograde CSD size is constrained \citep{Martin2011} by $R_H$, while even high turbulence $\langle \alpha_R \rangle \sim 0.1$ is unable to disrupt the classical prograde flow pattern around EBHs with circular orbits. 
This result may be expected since the typical turbulence scale $H$ is now smaller than the CSD size, 
{and suggests that transition from orderly to stochastic accretion may have a sensitive mass dependence \footnote{We note that according to the argument of \citet{McKernan2022}, 
if the spin of high mass EBHs dominating $\chi_{\rm eff}$ in extremely unequal mass-ratio mergers tend to be more systematically aligned with the global disc, 
they can contribute to a large $
\chi_{\rm eff}$ feature for unequal mass-ratio GW events consistent with observations \citep{Callister2021}.}. 
Relaxing the degrees of freedom in 3D simulations might also change the quantitative picture.
In this work dedicated to laying out an analytical framework for long-term chaotic spin growth, 
we simply limit our discussion to sub-thermal companions around high-mass SMBHs and make the following assumptions based on our numerical results: }

{1) Under influence of strong turbulence, the CSD flow direction of sub-thermal EBHs can become chaotic, with each accretion episode lasting a typical timescale of $\Delta t_* \sim 2\pi/\Omega$. }

{2) Our simulation applies 2D geometry, while typical turbulent eddies generated by 
both gravito-turbulence and MRI becomes isotropic on scales $<H$ \citep{Beckwith2011,BoothClarke2019}. Taking this effect into consideration, we make an extrapolation to 3D. 
Instead of switching between prograde and retrograde with respect to the global disc rotation, the dominant eddy that becomes regulated into a CSD by the companion gravity in its vicinity has a fairly isotropic distribution of average inclination $\theta$ with respect to the current BH spin, 
introducing the necessity of including LT effect. However, 
in Figure \ref{fig:std} we will show that the magnitude of final spins is not sensitive to this extrapolation.}

\section{Long-term Evolution of EBH Spin Parameter}
\label{sec:individual_spin}
The final accretion rate onto the EBHs is likely constrained by the Eddington limit 
$\dot{M}_\filledstar =  L_E/\eta_\filledstar c^2$, where the EBH luminosity reaches its Eddington luminosity $L=L_{\rm E} = 1.25\times 10^{38} M_\filledstar/M_\odot 
\text{erg s}^{-1}$, and $\eta_\filledstar$ is the efficiency factor of EBH accretion.
{The EBH's mass-growth timescale is 
\begin{equation}   
\tau_{\rm M}  \approx \dfrac{M_{\filledstar}}{\dot{M}_\filledstar} \simeq  \eta_\filledstar \tau_{\rm Sal}
\end{equation}
where  $\tau_{\rm Sal} =M_\filledstar c^2/ {L}_{\rm E} = 4.5 \times 10^8 \ {\rm yr}$ is the Salpeter timescale.} Within an 
order-of-magnitude, gas in the circumstellar discs (CSDs) is accreted onto the EBHs at $R_{\rm isco}$
with specific angular momentum $|J_c({R}_{\rm isco})| \sim \sqrt{GM_{\filledstar}{R}_{\rm isco}}$, such that in a quiescent environment the spin parameter
$|a|$ evolves towards unity on a similar timescale as $\tau_M$.

The accretion efficiency $\eta_\filledstar$, which also depends on black hole mass and spin, is usually on the order of a few percent for isolated black holes \citep{Jiang2014,Jiang2019} or even lower due to possible strong outflow/jet for the EBHs in AGN discs \citep{Tagawa2022}, which may provide an important source of heating to the global disc environment. Here we neglect feedback effects as in our simulations, and assume $\eta_\filledstar$ to be a constant such that $\tau_M$ can be a natural unit in our calculations, and 
$M_{\filledstar}(t) = M_{\filledstar}(t=0)\exp{(t/\tau_M)} $
is a universal mapping of how $M_{\filledstar}$ evolves with time.

Generally in a turbulent medium, gas accretion occurs in randomly oriented episodes. For SMBH growth over cosmic time, the duration of accretion episodes may be characterised by the timescale that a total ``self-gravitating-disc" mass is accreted at the Eddington rate \citep{King2006,king2008}. {Alternatively, applied to stellar-mass EBHs in 
an AGN disc, our simulation shows that the episode timescale is comparable to local dynamical timescale $\Delta t_\filledstar \simeq 
2\pi/\Omega \ll \tau_M$, which generally reflects the eddy-turnover or auto-correlation time for MRI and gravito-turbulence 
\citep{Oishi2007,BaruteauLin,BoothClarke2019}. We consider the appropriate cadence limit such that the number of spin-reorientation episodes during one accretion timescale $\tau_{M}$ is}
\begin{equation}
 \mathcal{S} \equiv \tau_{M}/\Delta t_\filledstar \simeq  \tau_{M}\Omega/2\pi  = 3\times 
  10^4 \dfrac{\eta_\filledstar}{0.1} \left(\dfrac{M_{\bullet}}
  {10^8 M_\odot} \right)^{1/2} \left(\dfrac{r_0}{\rm 0.3 pc} \right)^{-3/2} 
\end{equation}

Since the EBH's $\eta_\filledstar$ does not necessarily equal to the \textit{SMBH's} accretion 
efficiency $\eta_{\bullet}$, there are $\mathcal{S} \eta_{\bullet}/\eta_\filledstar$ cycles within the \textit{SMBH's} growth timescale or the AGN lifetime. But here we consider $\eta_{\bullet} \sim \eta_\filledstar$ such that the AGN lifetime is comparable to $\tau_M$ of individual EBHs.
Given the ratio $\Delta t_\filledstar/\tau_M = \mathcal{S}^{-1}$ as the frequency parameter, we numerically model the evolution of EBH spin as a function of time and mass by the following procedure.

1) During one single continuous accretion episode, starting with an initial black hole mass $M_0$ and initial $a_0$, the initial normalized value of
$\mathcal{R}_{\rm isco}= {R}_{\rm isco}/R_{\filledstar}$ ($R_{\filledstar}$ is the Schwarzschild radius) can be calculated from the generic relation between $\mathcal{R}_{\rm isco}$ and $a$ \citep{Bardeen1970,
Tagawa2020,Reynolds2021}:

\begin{equation}
    \mathcal{R}_{\rm isco} = 3+Z_{2} - \text{sign} (a) \sqrt{\left(3-Z_{1}\right)\left(3+Z_{1}+2 Z_{2}\right)}
    \label{eq:risco}
\end{equation}

\begin{equation}
    \begin{gathered}
Z_{1}=1+\left(1-|a|^{2}\right)^{1 / 3}\left[(1+|a|)^{1 / 3}+(1-|a|)^{1 / 3}\right] \\
Z_{2}=\sqrt{3|a|^{2}+Z_{1}^{2}}
\end{gathered}
\label{Z1Z2}
\end{equation}

The magnitude of
$\mathcal{R}_{\rm isco}$ ranges from 9 at $a=-1$,
to $6$ at $a=0$, then $\sim 1$ as $a$ approaches unity. 
In the absence of any  discontinuous change in the CSD spin direction and $\mathcal{R}_{\rm isco}$, 
the quantity $\mathcal{R}_{\rm isco}^{1/2}M_{\filledstar} = \mathcal{R}_{\rm isco}^{1/2}(M_0, a_0) M_{0}:= \mathcal{R}_{\rm isco,0}^{1/2}M_{0}$ is conserved \citep{Bardeen1970,King2006}, and $a$ evolves as a mapping of $M_\filledstar$:

\begin{equation}
\begin{aligned}
    a(M_{\filledstar})  = \dfrac13 \mathcal{R}_{\rm isco}^{1/2} \dfrac{M_0}{M_{\filledstar}}\left[4-\left(3\mathcal{R}_{\rm isco,0} 
    \left(\dfrac{M_0}{M_{\filledstar}} \right)^2-2\right)^{1/2} \right]
    \label{eqn:evolvea}
\end{aligned}
\end{equation}


2) In these classical equations, the sign of $a$ is determined by the directions of black hole angular momentum vector ${\mathbf{J}}_{\filledstar}$ and local CSD angular momentum $ {\mathbf{J}}_d$ (specific magnitude of which becomes relevant in 3D turbulence), such that $a := |a|\text{sign} ({\mathbf{J}}_{\filledstar}\cdot {\mathbf{J}}_d)$ \citep{Tagawa2020,Reynolds2021}. But if the direction of ${\mathbf{J}}_d$ of a population of EBHs changes intermittently due to fluctuating turbulence, this definition of $a$ is unimportant in a collective sense. In our formulation, there is an absolute vertical direction $\hat{\mathbf{z}}$ associated with the global SMBH accretion disc's prograde direction, and 
only the absolute spin $ |a|\text{sign} ({\mathbf{J}}_{\filledstar}\cdot\hat{\mathbf{z}})$ may change continuously between switching of EBH accretion cycles, while $a$ alternates between positive and negative values, and $\mathcal{R}_{\rm isco}$ changes discontinuously (Equation. \ref{eq:risco}) between values larger and smaller than 6. 

\begin{figure*}
\centering
\includegraphics[width=0.48\textwidth]{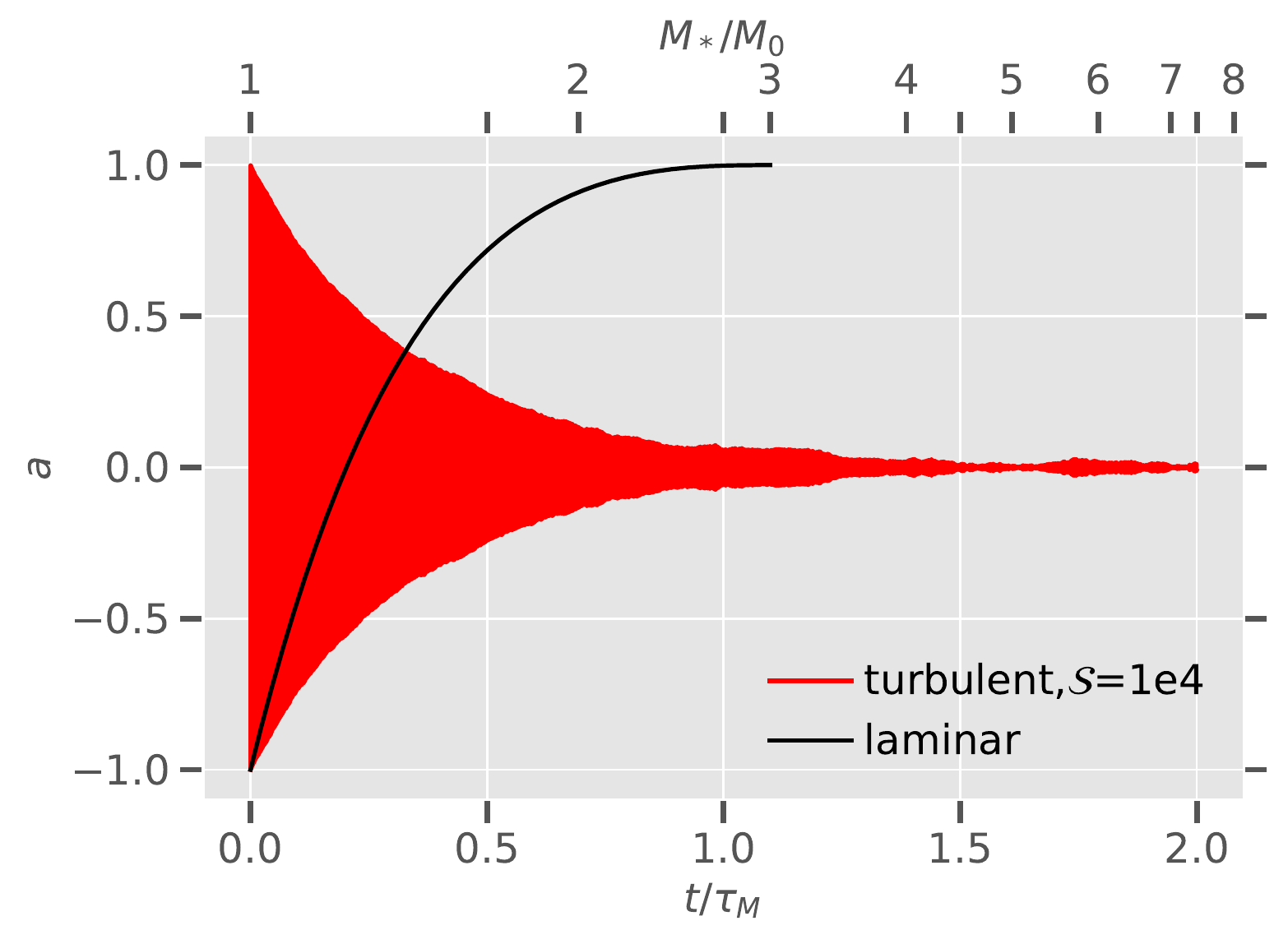}
\includegraphics[width=0.48\textwidth]{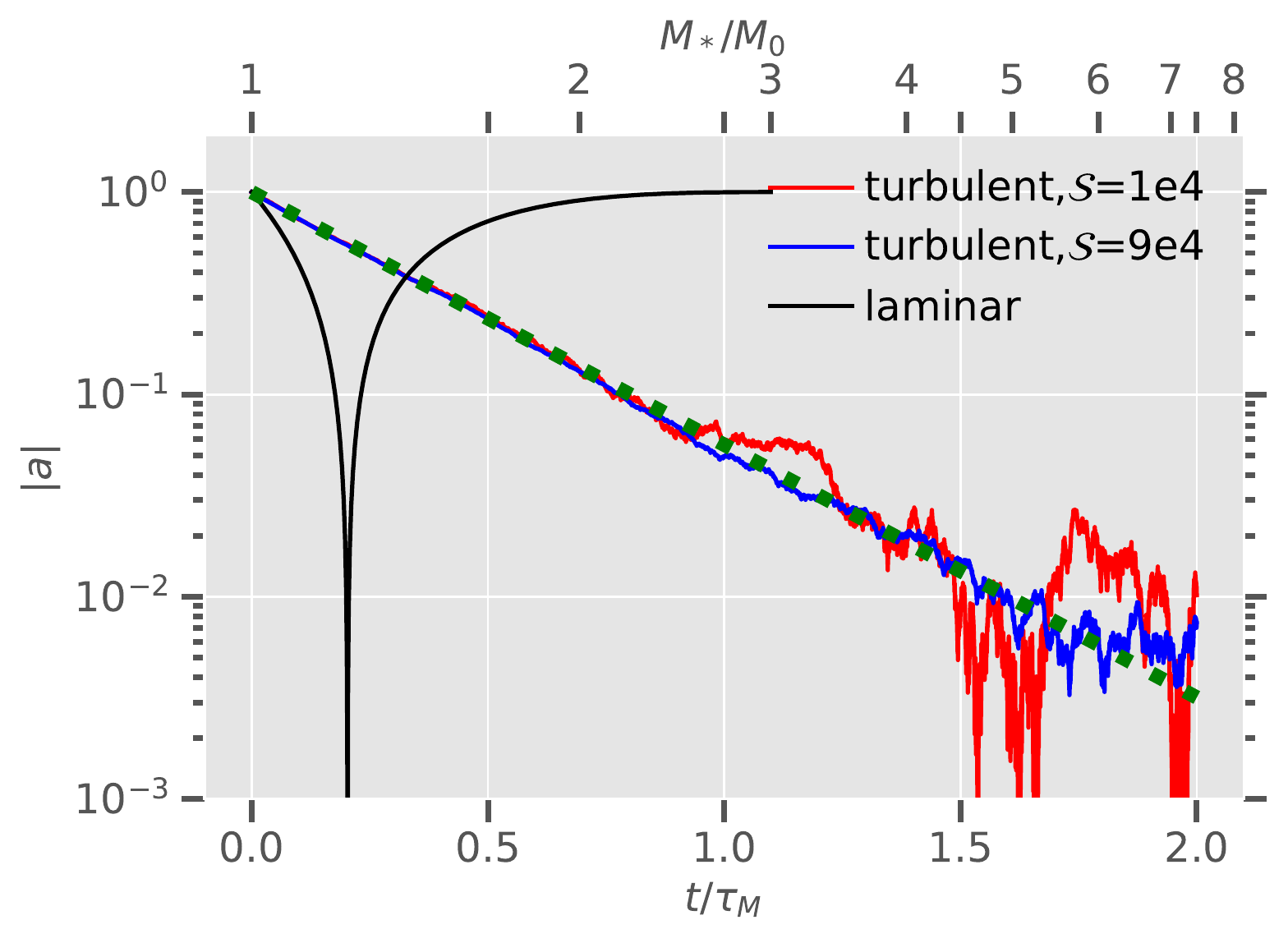}
\caption{Left panel: Evolution of EBH spin $a$ with sign determined with respect to CSD spin for the laminar case (fixed CSD spin direction, black line), and turbulent case with $\mathcal{S}=10^4$ (red line). In the turbulent case, the CSD spin changes discontinuously every accretion episode so $a$ shifts between positive and negative values with a trend of decaying in magnitude; Right panel: Evolution of $|a|$ in these cases, with an additional $\mathcal{S}=9\times 10^4$ case shown in blue solid line. The green dotted line shows analytical prediction for the initial spin-down phase, before the spin magnitude becomes comparable to $\mathcal{S}^{-0.5}$.}
\label{fig:individual}
\end{figure*}

The left panel of Figure \ref{fig:individual} shows two exemplary individual cases of EBH spin $a$ evolution. By convention, the sign of $a$ 
is determined {\it relative to the local CSD flow} 
\citep{Bardeen1970,Tagawa2020}, or explicitly, $a =  |a| 
\text{sign}(\cos\theta) = |a| \text{sign}(\mathbf{{J_d} \cdot {J_\filledstar}}) $. 
In a laminar AGN disc, 
a steady-state CSD flow is either prograde or retrograde with respect to the absolute global disc rotation \citep{liYP2022,chen2022prograde}, 
and EBH spin monotonically grow towards this preferred direction. 
The solid black line shows how an initially 
counter-aligned $a=-1$ EBH would grow its spin towards the CSD rotation axis, with mass increasing exponentially on 
the doubling timescale $\tau_{\rm M} = \tau_{\rm Sal}\eta_\filledstar$ through Eddington-limited accretion, where $\tau_{\rm Sal}$ is the Salpeter timescale and $\eta_\filledstar$ is the EBH accretion efficiency. 
In this ideal reference
case, we assume initial spin axis  is counter-aligned with CSD spin, and in time $\theta = \pi$ 
would  discontinuously jump to $\theta = 0$ as $a$ crosses over to positive, since for $\theta = \pi,0$ there 
is no Lense-Thirring (LT) precession torque to change $\theta$ 
continuously. 
On this ``fundamental track", $\mathcal{R}_{\rm isco}$ shrinks by a factor of 9 as $M_\filledstar$ grows by a factor of 3 from its initial value, reaching an asymptotic limit of $a\simeq 1$ 
after a timescale of $\tau_{\rm M}\ln(3)$. 
EBHs born with spin larger than -1 starts somewhere middle on 
this same track but the final spins all converge towards 1.

The red line in left panel of Figure \ref{fig:individual} shows evolution of $a$ in a fiducial turbulent case with  $\mathcal{S}=10^4$. 
The initial magnitude of the spin is $|a|=1$, but at the start of every accretion episode with constant 
duration $\tau_{\rm M}/\mathcal{S} = \tau_{\rm Sal}\eta_\filledstar /\mathcal{S}$, the orientation of 
CSD spin is randomized with respect to the current EBH spin, and we take account of LT torque in the 
evolution of $\theta$ during every episode. Although $a$, by definition, oscillates 
between positive and negative values due to the sporadic shift of CSD spin, through plotting the 
evolution of $|a|$ in the right panel of Figure \ref{fig:individual} in logarithmic scale, we can more clearly see a 
continuous change in the spin magnitude. We found in the turbulent case, the typical value of $|a|$ 
first evolves as an exponential decay due to the intrinsic asymmetry between spin-up and spin-down, but 
after some typical spin-down timescale $\tau_d$ comparable to $\tau_{\rm M}$, random fluctuations around
$a=0$ due to accretion of individual cycles dominate, and characteristic values fluctuate around 
the random-walk dispersion $a_{RW}\sim \mathcal{S}^{-0.5}$. 
We also tested with $\mathcal{S}=9\times 10^4$ 
(right panel, blue line), in which case the random walk factor is smaller, and a longer decay timescale 
$\tau_d$ is needed for the initial spin to decay below this factor. The green dotted line shows 
analytical prediction $|a|\approx \exp (- 2.876 t 
/ \tau_{\rm M})$, which describes very well  the initial 
spin-down phase. The derivation of this prescription as well as expression
$\tau_d \approx \tau_M\ln \mathcal{S}/5.752$ are elaborated in Appendix \ref{app:spin_decay}.

\section{Evolution of Spin Orientation Due to Turbulence \& Lense-Thirring Torque}
\label{sec:Lense-Thirring}

If the effective disc angular momentum that exerts the LT torque is much smaller than the BH angular momentum, 
the CSD will generally evolve towards alignment with the EBH if $\theta <\pi/2$ and counteralignment if $\theta > \pi/2$, on a LT timescale of

\begin{equation}
\tau_{\rm LT}  = \dfrac{J_\filledstar}{J_d/\Delta t} \simeq {|a| G M_\filledstar^2/c \over  ( L_E/\eta c^2)  \sqrt {G M_\filledstar R_{\rm w}}} 
\simeq { R_\filledstar^{1/2} \over R_{\rm w}^{1/2}} |a| \tau_{\rm M},
\label{eq:tautl}
\end{equation}
the warp radius is given by \cite{King2005} 

\begin{equation}
    \begin{aligned}
\frac{R_{\mathrm{w}}}{R_{\filledstar}}=& 990\left(\frac{\eta}{0.1}\right)^{1 / 4}\left(\frac{L}{0.1 L_{\mathrm{E}}}\right)^{-1 / 4} \left(\dfrac{M_\filledstar}{10^8 M_\odot} \right)^{1 / 8} \\
& \times\left(\frac{\alpha_{1}}{0.03}\right)^{1 / 8}\left(\frac{\alpha_{2}}{0.03}\right)^{-5 / 8} |a|^{5 / 8}\\ & \approx 100 |a|^{5/8} := \mathcal{R}_{\rm w,0} |a|^{5/8}
\end{aligned}
\end{equation}

where $\alpha_1, \alpha_2$ are the accretion and warp-propagation viscosities. We assume the typical 
value of $\alpha_1 \sim \alpha_2\sim 0.03$ \footnote{applying $\alpha_1 \gtrsim 0.1$ directly from our numerical simulation makes little difference}, and choose $\mathcal{R}_{\rm w,0}=100$ relevant to our EBH parameter 
$M_\filledstar \sim 10-100M_\odot$. We also limit $R_w/R_{\filledstar}>1$. In our case we have defined $J_d = (\eta L_E/c^2)  \sqrt {G M_\filledstar R_{\rm w}} \Delta 
t_\filledstar$ as the angular momentum that flows past the warp radius \citep{king2008} during one 
accretion cycle. Note that $J_d = J_c(R_w) \dot{M} \Delta t_\filledstar$ is generally much larger 
than $J_c({R}_{\rm isco}) \dot{M} \Delta t_\filledstar$: the former is the total angular momentum
responsible for exerting the LT torque, and the latter is only its small fraction 
that gets directly fed onto the black hole through ${R}_{\rm isco}$.

When $\Delta t_\filledstar/\tau_M$ is very small or $\mathcal{S}$ is large, generally $J_d \ll J_{\filledstar}$ and $\tau_{LT}\gg \Delta t_\filledstar$ for moderate values of $|a|$, which means LT torque cannot strongly influence the EBH inclination during any short accretion cycle, and while $\theta$ changes randomly between accretion cycles due to jumps in CSD spin axes, the LT effect cannot accumulate in any preferred direction, so the evolution of $a$ is nearly independent of $\theta$. However, ${J_{d}}/{J_{\filledstar}} \propto |a|^{-11/16} $ increases with a decreasing $|a|$, and $J_d = J_\filledstar$  when
$\tau_{LT} = \Delta t_\filledstar$ or $  a_{\rm crit} \simeq (R_w(a_{\rm crit})/R_{\filledstar})^{1/2} /\mathcal{S}$,
which gives
\begin{equation}
    a_{\rm crit} \simeq (\mathcal{R}_{\rm w,0}/{\mathcal{S}}^2)^{8/11} .
    \label{eqn:acrit}
\end{equation}

For $|a|\gtrsim a_{\rm crit}$, any initial $\theta > \pi/2$ is generally directed towards $\pi$ (counter-alignment) by LT torque, but for $|a|\lesssim a_{crit}$, even $\theta > \pi/2$  might be directed towards $\theta = 0$ on a timescale of $\tau_{LT} < \Delta t_*$, which leads to a systematic spin-up of the magnitude of $|a|$ until it fluctuates around the quasi-steady value of $\pm a_{\rm crit}$ \footnote{The exact long-term alignment criterion is $-2 J_{\rm BH}\cos (\theta) < J_d$ \citep{King2005}, therefore technically $ J_{\rm BH}<J_d/2$ is needed to guarantee systematic spin-up for any random $\theta$, but extra order-unity factors do not qualitatively affect our argument}. 
Thus concluded \citet{King2006} in their qualitative analysis relevant for SMBHs, but they did not consider influence of the random walk. We 
demonstrated in Figures \ref{fig:pdf} \& \ref{fig:std} that the distribution of $|a|$ is dominated by random walk when $1/\sqrt{\mathcal{S}} \gg  a_{crit}$, but will show later that we can reduce to their scenario when $1/\sqrt{\mathcal{S}} \lesssim a_{crit}$ which is probable for SMBH growth, although unlikely in our context.

In our turbulent models for EBH spin evolution, at the start of every turbulent episode, we pick the initial inclination of $\mathbf{J_{\filledstar}}$ with respect to $\mathbf{J_{d}}$ from a uniform isotropic distribution, equivalent to picking $\cos (\theta)$ from a uniform distribution from -1 to 1. 
If $\cos(\theta)$ has changed sign compared to the previous cycle, $a$ would also change sign and we discontinuously update $\mathcal{R}_{\rm isco}$ with Equation \ref{eq:risco}, and then evolve $a$, $\mathcal{R}_{\rm isco}$ continuously again with Equation \ref{eqn:evolvea} (replacing $\mathcal{R}_{\rm isco,0}^{1/2}M_0$ with the updated $\mathcal{R}_{\rm isco}^{1/2}M_\filledstar$), until $\cos(\theta)$ changes sign again either due to LT or continuous spin accretion. 
The short-term continuous local evolution of $\theta$ during every accretion timescale of $\Delta t$ can be calculated by \citep{King2005}

\begin{equation}
    \dfrac{d}{dt}\cos\theta \approx  \dfrac1{\tau_{\rm LT}J_{BH}} \sin^2 \theta (J_d + J_{BH}\cos \theta),
    \label{thetaevolution}
\end{equation}

where $\tau_{\rm LT}$ is used as a normalization for the dissipation term. Note that $\tau_{\rm LT}$ is only an estimate of the alignment timescale for moderate values of $\theta$, when $\theta \approx \pi$ the actual timescale becomes much larger than $ \tau_{LT}$ and approaches infinity at $\theta =\pi$ even if $J_\mathrm{BH} \ll J_d/2$ since there is no LT precession. 


When $\theta < \pi/2$, the time derivative of $\cos(\theta)$ is always larger than 0 and $\theta$ decreases towards 0 ($|a|$ consistently spins up). When $\theta > \pi/2$, however, it is worth clarifying that the long-term evolution of $\theta$ is not immediately obvious from its local evolution. For example, if  $J_{BH} > J_d$ and $\theta \gtrsim \pi/2$ we should have long term counteralignment, but if the initial $\cos (\theta)$ during this cycle is infinitely close to zero or more generally roughly corresponds to a range of $-J_d/J_\filledstar \lesssim \cos\theta \lesssim 0$, its derivative is actually positive, which would momentarily align the BH with the disc ($|a|$ spins up
momentarily). When that happens, we also update $a$ and $\mathcal{R}_{\rm isco}$ discontinuously \textit{during} an accretion cycle before applying Equation \ref{eqn:evolvea}. 
This prescription does not contradict the long-term counter-alignment criterion since over a longer timescale $\sim \tau_{LT}$ the EBH would eventually tilt back to become counter-aligned with $J_d$ on timescales comparable to $\tau_{LT}$, see Figure 2 of \citet{King2005}. However, practically when $\Delta t_\filledstar < \tau_{LT}$, the EBH may not be able to counter-align again before another \textit{new} accretion cycle kicks in and $\cos(\theta)$ is randomized again, so the spin-down is indeed changed to spin-up midway through an accretion cycle even when $J_{BH} > J_d$ in these ``lucky" cases. 
In the ``lucky" cases, the EBH spin is momentarily aligned with the disc, may not grow back towards long-term counter-alignment before the next accretion cycle cuts in.

Nevertheless, our result shows that $|a|$ still relaxes towards typical values around $a_{RW} = 1.5\mathcal{S}^{-1/2}$, and on average $|a|$ never gets below $a_{\rm crit}\ll a_{RW}$ for LT torque to have a strong effect and for the eternal-alignment criterion to play a role. This is because the ``lucky" cases roughly corresponds to a range of $-J_d/J_\filledstar \lesssim \cos\theta \lesssim 0$, the chance of which happening for isotropic $-1 < \cos\theta < 1$ during every accretion episode is on the order of

\begin{equation}
    \left( \dfrac{J_d}{J_\filledstar} \right) \sim \left(\dfrac{a_{\rm crit}}{|a|}\right)^{11/16},
\end{equation}

which self-consistently is much smaller than order-unity when $|a|$ stabilizes around the the random walk equilibrium $a_{RW} =  1.5 \mathcal{S}^{-1/2} \gg a_{\rm crit}$, and the general evolution of $|a|$ still turns out to be random-walk dominated.

By comparing $a_{RW}$ with Equation \ref{eqn:acrit} which shows $a_{\rm crit}$ as a steeper power law of $\mathcal{S}$, we have

\begin{equation}
    \dfrac{a_{\rm crit}}{a_{RW}} \approx 20 \left(\dfrac{\mathcal{R}_{w,0}}{10^2}\right)^{8/11} \mathcal{S}^{-21/22}.
\end{equation}

This equation implies that when $\mathcal{R}_{w,0}\gtrsim 100$ and $\mathcal{S} \lesssim 20$, $a_{\rm crit}$ may still become larger than $a_{RW}$. But this range of $\mathcal{S}$ is too small to be relevant in the current context, i.e. $\Delta t_\filledstar \simeq 2\pi/\Omega$. 

\section{The Population Model, with and without Lense Thirring Effect}
\label{sec:pop}
In the population models, given the re-orientation parameter $\mathcal{S}$, the spin of $10^3$ EBHs are evolved over a timescale of $3\tau_M$, with each initial spin $a_0$ sampled from a uniform initial distribution from -1 to 1. 
The initial orientation is also randomly chosen. 
The initial mass function is irrelevant to the spin evolution in our setup.

\begin{figure*}
\centering
\includegraphics[width=0.5\textwidth]{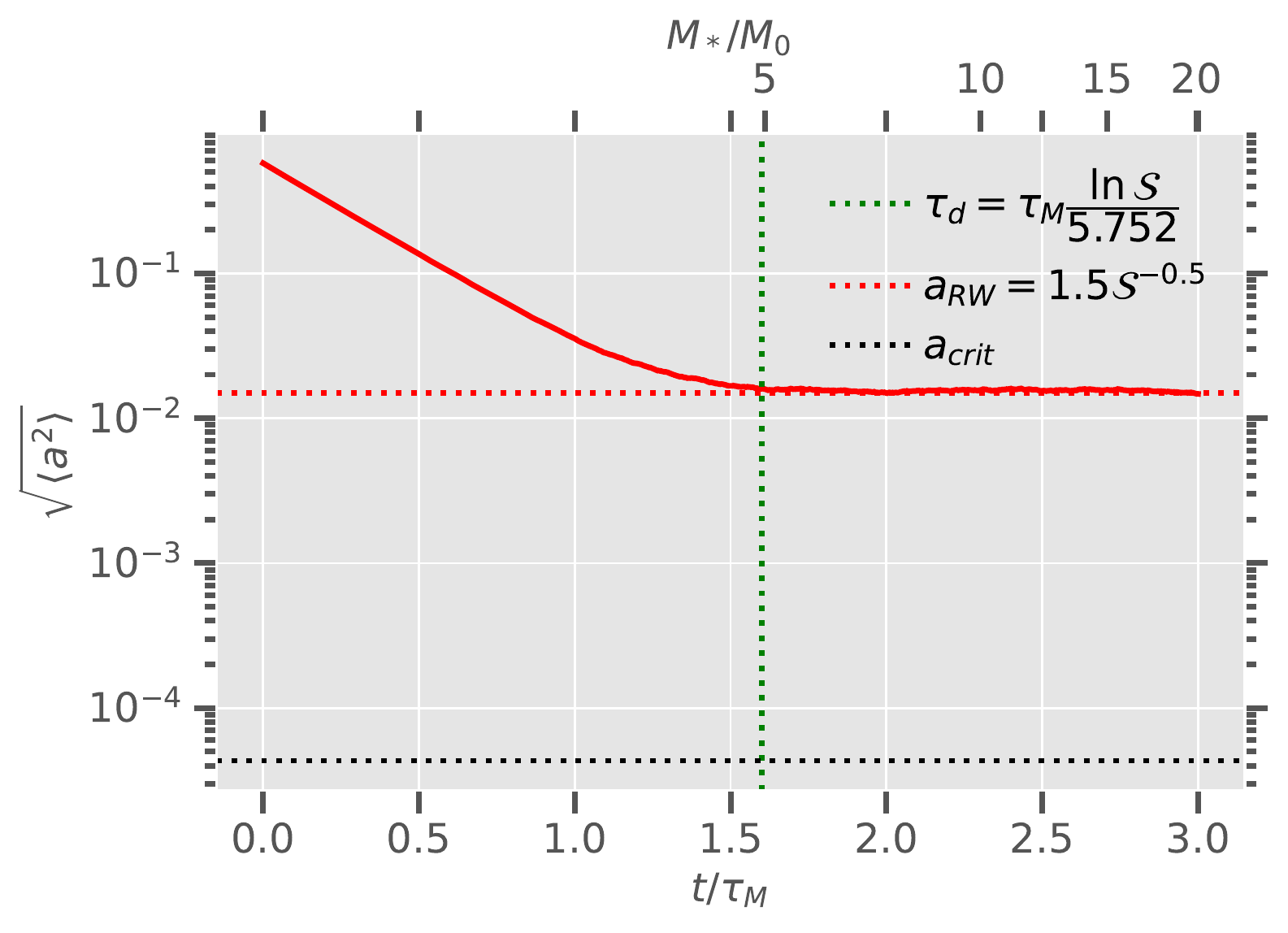}
\includegraphics[width=0.47\textwidth]{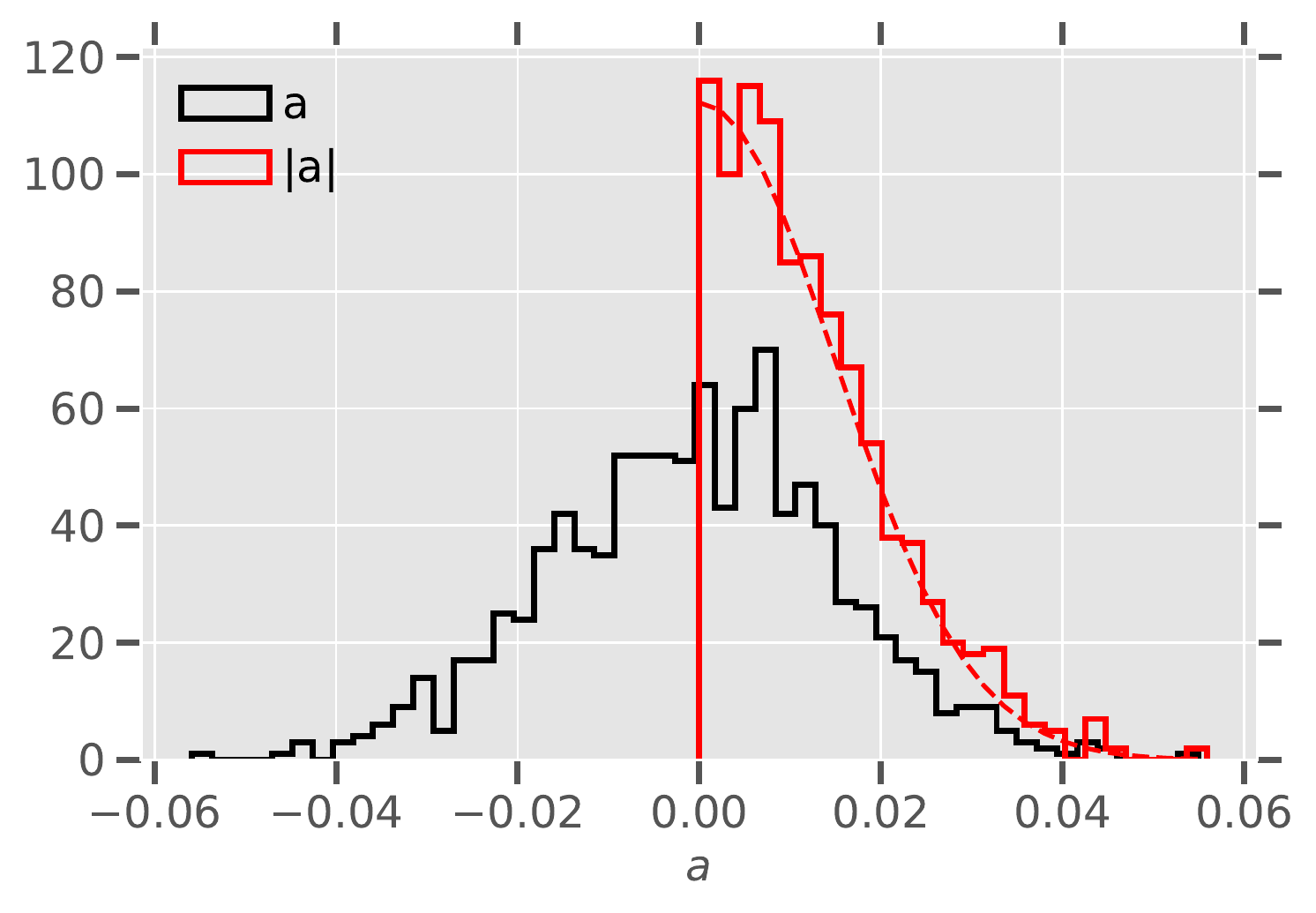}
\caption{Spin evolution of a population of $10^3$ EBHs with $\mathcal{S}=10^4$. Left panel: the root-mean-square of spin parameter, evolving with time, shown in red solid line. The decay time (green dotted line) serves as an accurate estimate of the time where initial $\sqrt{\langle a^2 \rangle}=1/\sqrt{3}$ (for uniform distribution) decays to the random walk factor $a_{RW}$ (red dotted line), which is much larger than the critical spin (black dotted line) required for systematic spin-up by LT torque; Right panel: The histogram of $a$ and $|a|$ after 3 $\tau_{\rm 
M}$, reaching a semi-steady state. The distribution of $|a|$ can very well be approximated by half a Gaussian with standard deviation $a_{RW}$. 
}
\label{fig:pdf}
\end{figure*}

\begin{figure}
\centering
\includegraphics[width=0.5\textwidth]{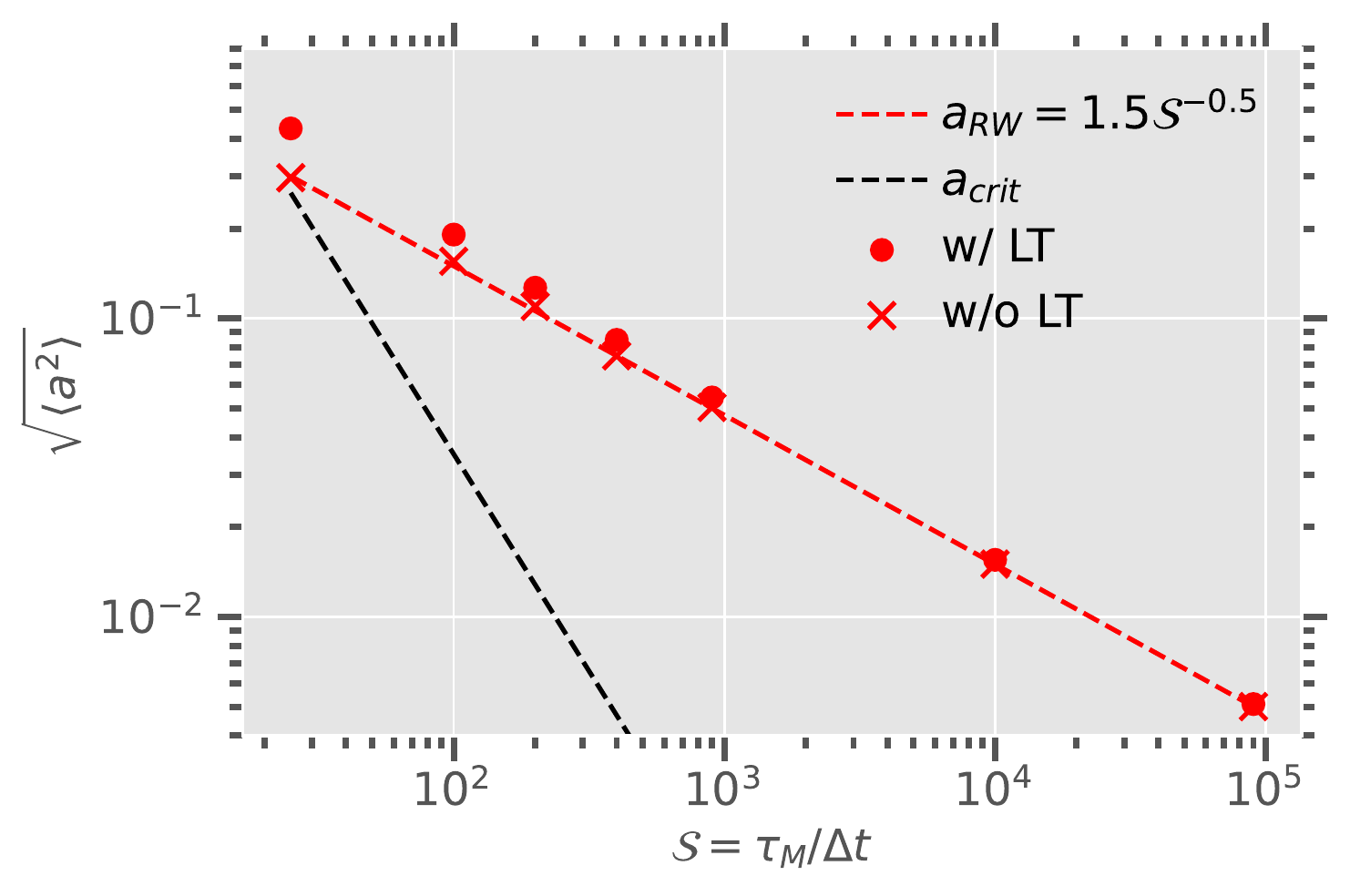}
\caption{Red dots: The steady-state root mean square spin value of EBH populations after $3\tau_M$ of evolution, varying the spin-reorientation number $\mathcal{S}$. $a_{RW}$ as a function of $\mathcal{S}$ is shown in red 
dashed line and $a_{\rm crit}$ in black dashed line. For large $\mathcal{S}$ and $a_{crit}\ll a_{RW} $ we verify that the dispersion converges to $a_{RW}$, and the $|a|$ histograms are consistently Gaussian. when $a_{\rm crit}, 
a_{RW}$ becomes comparable, the final dispersion begins to deviate from $a_{RW}$ due to LT effects. Nevertheless, $\mathcal{S}\lesssim 100$ is highly unrealistic for the dynamical timescale of disc turbulence and EBH masses.}
\label{fig:std}
\end{figure}

In the left panel of Figure \ref{fig:pdf} we show evolution of average dispersion $\langle a^2 \rangle$ for a population of
$10^3$ EBHs, 
starting from a uniform distribution of $a$ between -1 and 1. 
The root mean square of the spin converges towards an asymptotic value of $a_{RW} \approx 1.5 
{\mathcal{S}^{-0.5}}$ (red dotted horizontal line), and it does not grow subsequently 
as it would have in a pure RW. This outcome is due to the competing effect of 
spin down and RW. With $|a| \approx a_{RW}$, the spin down effect that reduces $|a|$ has 
stricken an equilibrium with RW diffusion that tends to expand $|a|$. The decay 
time $\tau_d$ is shown as the green dotted line, which serves as an estimate of 
the time of transition from initial spin-down-dominated phase to a RW-dominated phase 
on a population level, since it reflects the slowest possible spin-down timescale of 
any EBH within this population. The critical spin $a_{crit}$ (black dotted line) below which LT effect 
becomes important is a sensitive function of $\mathcal{S}$ (see Figure \ref{fig:std}). 
Since the characteristic value of $|a|$ never gets below $a_{RW}\gg a_{\rm crit}$, LT 
torque does not play a significant role in this evolution process.

In the right panel of Figure \ref{fig:pdf}, we show the histogram for $a$ and $|a|$. Note that due to the slight asymmetry between spin up and spin down with respect to $\mathbf{J_d}$ (spin down is more efficient), the PDF of $a$ has a mean value slightly shifted towards the positive. However by formulation, the $\mathbf{J_d}$ vector distribution is also isotropic, therefore the distribution of spin \textit{projection} $|a| \text{sign} (\mathbf{J_d \cdot z})$ with respect to any reference absolute vector $\mathbf{z}$ would essentially be a symmetrically expanded version of the magnitude $|a|$ distribution. We show that the $|a|$ histogram can be approximated very well by Gaussians with dispersion $1.5 \mathcal{S}^{-0.5}$ (dashed red line).

{To illustrate the relative importance of LT torque, we also run population evolution of
EBHs without LT effect, in which we randomly update $\cos(\theta) = 1$ or $\cos(\theta) = -1$ at the beginning of each spin-reorientation
episode but do not allow it to evolve, artificially confining the CSD to be either prograde or retrograde with respect to the global disc as in our 2D geometry simulations, just for numerical comparison.


Nevertheless, in the relevant parameter space, the spin evolution is expected to be dominated by random walk (Figure \ref{fig:std}, red crosses) and not much different from the 3D model.
}

\begin{figure}
\centering
\includegraphics[width=0.48\textwidth]{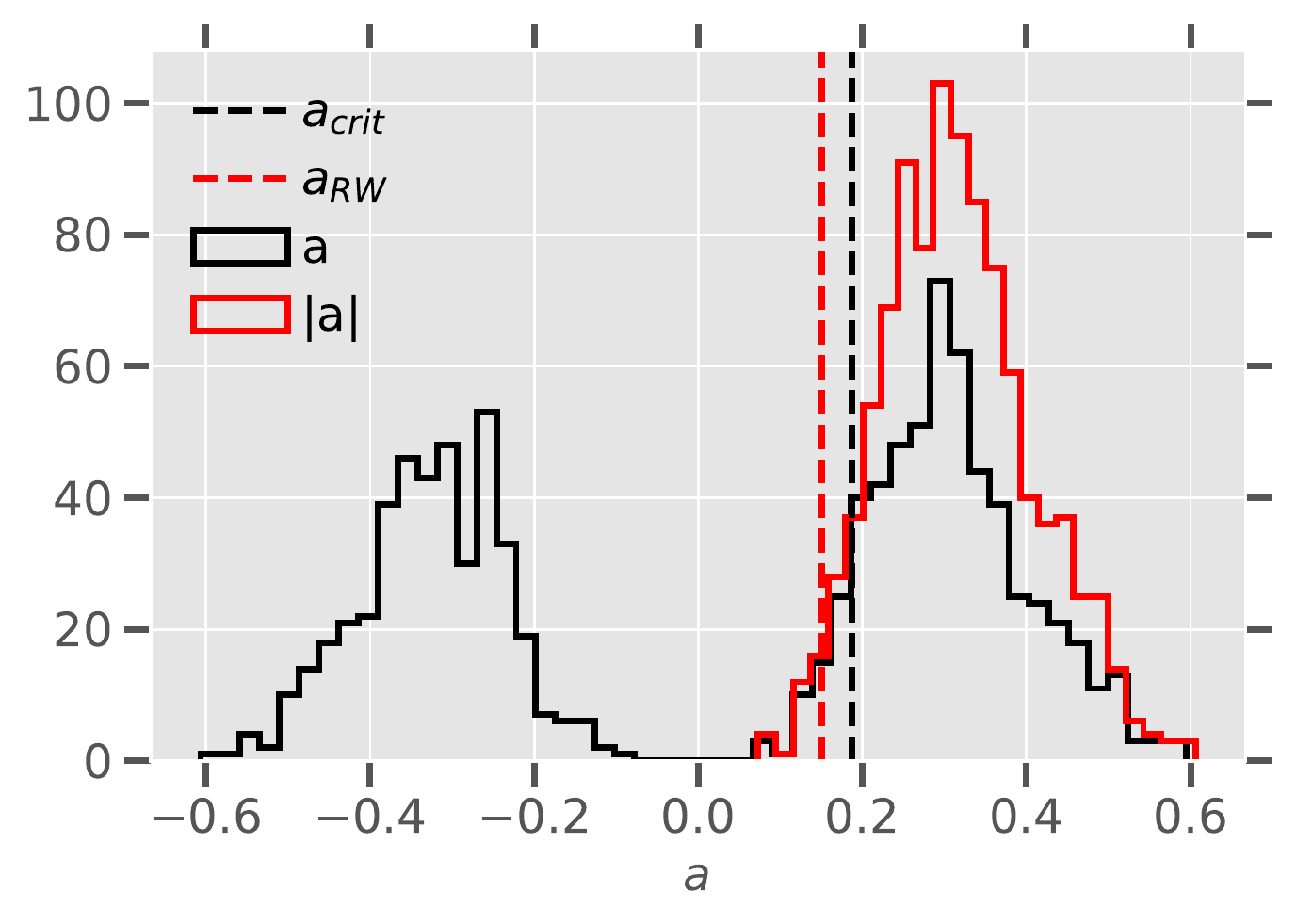}
\caption{The semi-steady distribution of $a$ and $|a|$ after 3 $\tau_M$ of evolution for the SMBH context, with $\mathcal{R}_{w,0}=1000, \mathcal{S}=100$. The distribution of $|a|$ can no longer be approximated by Gaussian, but is strongly peaked around values close to $a_{\rm crit}$ (dashed black line), which is already dominating the random walk factor (red dashed line), since systematic spin-up by LT torque can clear out a deficit for any $|a|<a_{\rm crit}$. It can be compared to Figure 4 of \citet{king2008}.}
\label{fig:PDF100SMBH}
\end{figure}

As a sanity-check, we demonstrate explicitly that LT torque may still play a large role when a large $\mathcal{R}_{w,0}$ relaxes the viable range of $\mathcal{S}$ for $a_{\rm crit} > a_{RW}$, conforming with previous numerical simulations of SMBH growth \citep{king2008}, first qualitatively suggested \citet{King2006}. Considering the central SMBH mass $M_{\bullet}\sim 10^8 M_\odot$, we can choose $\mathcal{R}_{w,0}=1000$ which is 
an order of magnitude larger than that in the EBH context. 
The accretion timescale of \citet{king2008} is defined in terms of the self-gravitational disc mass $\tau_{sg}\sim 0.01 \tau_M$, which is similar to 
$\mathcal{S} \simeq 10^2$. 
For these parameters, we confirm that $a_{\rm crit}>a_{RW}$.
The corresponding $a$ histogram for a population of $10^3$ SMBHs after $3\tau_M$ of evolution is shown in Figure \ref{fig:PDF100SMBH}. In this case, the systematic spin-up by LT torque can prevent any spin-down below $|a|<a_{\rm crit}$, and clear out a central
deficit in the distribution function. The spin parameters are strongly peaked around values comparable to $a_{\rm crit}$, corresponding to equal spin-up and spin-down efficiency \citep{king2008}.

Nevertheless, we emphasize again that this skewed distribution is hard to achieve in our EBH context unless $\alpha_2$ is very small, especially when $\alpha_2/\alpha_1=2\left(1+7 \alpha_1^{2}\right) /\left[\alpha_1^{2}\left(4+\alpha_1^{2}\right)\right]$ is usually much larger than order unity \citep{ogilvie1999}. 
We conclude that in most of the stellar-mass EBH cases RW would dominate, in the sense when $a_{RW} \gg a_{\rm crit}$, the deficit for $|a| < a_{\rm crit}$, although it exists, is negligible in the entire Gaussian distribution.

\section{Conclusions}
\label{sec:conclusion}
{Results of our hydrodynamical simulations imply that in the limit that the EBH mass is quite small ($\lesssim 100 M_\odot$) compared to the SMBH mass ($\gtrsim 10^8M_\odot$), 
under influence of strong fluctuating turbulence,} the inclination $\theta$ between CSD spin vector $\mathbf{J_d}$ and EBH spin vector $\mathbf{J_\filledstar}$ may be frequently
randomized, and the growth of EBH's spin occurs through a series of short and independent accretion episodes. 
The final characteristic 
magnitude of EBHs' spin is limited by a random walk (RW) factor on the order of $a_{RW} \sim \mathcal{S}^{-0.5}$, 
where $\mathcal{S}(\gg 1)$ is the number of CSD spin-reorientation episodes during each Eddington 
mass doubling timescale. 
Moreover, the spin axes distribution also becomes isotropic. 
The characteristic turbulence-coherence (eddy-turnover) timescale is on the order of the local orbital timescale, which 
gives $\mathcal{S}\sim 10^4$. 
A final spin distribution with low magnitude and isotropic direction implies a general 
low-$\chi_{\rm eff}$ distribution in subsequent merger events.
{This outcome is analogous to the low-asymptotic planetary spin 
due to planetesimal accretion versus the fast spins due to giant impacts \citep{dones1993}.}

 We extrapolate the above analysis for stand-alone EBHs to the merging BBHs observed by 
LIGO-Virgo.  If these BBHs form with very close separation and evolve quickly towards coalescence (before the 
components' spins are significantly modified by three-body encounters and/or multiple-disc interaction), the relative contribution of each member to the merger's $\chi_{\rm eff}$  generally cannot exceed the magnitude of $a_{\rm RW}$ of stand-alone EBHs (see Appendix \ref{app:chi_eff}).
In widely separated BBHs, the individuals' spins and the BBHs' orbital 
angular momentum may still have ample time to evolve and couple as they accret prior 
to merger. But, if they are surrounded by CSDs with persistent spin orientation,
non-negligible quadruple moment in the gravitational potential would induce precession
in their orbits. Moreover, either BBHs' modest migration through the global disc or their 
orbital contraction can lead to eccentricity and inclination excitation 
through evection and eviction resonances.  These effects significantly reduce the 
magnitude and reorient the direction of BBHs' orbital angular momentum \citep{2022arXiv220407282G}. Although these resonances might otherwise be suppressed for those BBHs embedded in CSD with 
frequent stochastic spin re-orientation (${\mathcal S} \gg 1$),  the cumulative consequence of 
accretion on the individual components' spins would be analogous 
to that of the stand-alone EBHs with asymptotic $\sqrt{\langle a^2\rangle} \sim a_{\rm RW}$, which sets an upper limit to $\chi_{\rm eff}$. An uncertainty in our model is the assumption
that that the turbulence field in the circum-SMBH is able to rapidly change the circumstellar flow
over an eddy-turnover timescale.  This assumption can be quantitatively tested with follow-up
numerical simulations.



During coalescence of EBHs which might contribute to some of the gravitational wave events detected by
LIGO, the EBH mass would increase monotonically with resulting $\vert a 
\vert \sim \mathcal{O} (1)$ \citep{hofmann2016}. Such growth may account for EBHs' larger masses in comparison
with the BHs in Galactic binary systems, but in order to reconcile with the low-$\chi_{\rm eff}$
found by LIGO, the coalesced EBHs need to substantially reduce their $\vert a \vert$ prior
to succeeding coalescence with other EBHs and/or randomize their spin axes. To achieve this, we suggest that EBHs' accretion of GI or MRI-induced
turbulent gas in the circum-SMBH discs can lead to both mass increases and $\vert a \vert$ decreases. 

The highly uncertain EBHs' merger timescale $\tau_{\rm merge}$
is determined by many effects including EBHs' migration, mass growth and BBHs' orbital evolution
under the influence of circum-BBH discs \citep{Li2021b,Li2021c,Li2022} and external secular perturbations \citep{2022arXiv220407282G}.  The efficiency of these competing mechanisms are beyond the scope of this paper, but we generally conclude that EBHs' mass increase may be primarily due to mergers/gas accretion in the limit $\tau_{\rm merge} \lessgtr \tau_{\rm M} =
\tau_{\rm Sal} \eta_\filledstar$ respectively.  The latter effect of diminishing $\vert a \vert$
is due to the ceaseless re-orientation of the relative angle between the EBHs' spin axis and the angular momentum of the turbulent gas accreted onto them. Considering the spin-down and RW-dominated phase of EBH spin evolution through gas accretion, in the limit $\tau_{\rm merge} \gtrsim \tau_{\rm M}$, the first phase takes about time 
$ \tau_d = \tau_{\rm M} \ln{\mathcal{S}}/5.752 \sim \tau_{\rm M}$ to erase any initial spin and the subsequent evolution
is dominated by random motion, until the dispersion reaches an asymptotic value $a_{RW} 
\sim \mathcal{S}^{-1/2}$ throughout the AGN duration. The
gas-accretion contribution would lead to the small $\chi_{\rm eff} 
(< 0.1)$ reported by the LIGO detection 
as well as a significant fraction of EBHs' mass growth. In the limit $\tau_{\rm merge}
\lesssim \tau_{\rm M}$ when merger is very frequent, the first spin-down phase can not be completed and
the lowest reachable value for the characteristic spin is 
$\sqrt{\langle a^2 \rangle} \sim \exp [- 2.876 \tau_{\rm merge} 
/ \tau_{\rm M}]$ before merger resets $a$ to $\mathcal{O}(1)$. 

{We conclude that spin-reorientation of CSDs, fed by rapidly-varying turbulent
global disc reconciles efficient gas accretion of EBHs with low spins, and reinforces the
scenario that AGN discs are  fertile hosting venues for BBH mergers. Furthermore, a low $\chi_{\rm eff}$ distribution from observation suggests that EBHs’ mass growth is dominated by gas
accretion rather than their coalescence and the energy dissipated during this 
process provides intense auxiliary heating sources for the global disc. }

{Finally, we remark that turbulent eddies can also stifle classical migration torques, limit and randomize radial migration of disc-embedded companions, although this global effect has  not been considered in this paper \citep{Laughlin2004,BaruteauLin}. The effect of disc turbulence on migration may help to obliterate the current outstanding population-synthesis issue of EBH mass distribution from AGN channel having typical mass that is too large compared with observation, by reducing the efficiency of EBH monotonically migrating into dense clusters around migration traps. While frequent dynamical interactions in migration traps was traditionally thought necessary to reduce the dispersion in $\chi_{\rm eff}$ \citep{Tagawa2020}, we have shown that turbulent accretion may be more efficient in producing low spin, which relaxes the need for dynamical interactions and in turn does not compulsorily lead to large EBH masses. Since turbulence may mutually constrain EBH spin and mass, this process is a 
relevant and central effect which should be incorporated in next-generation population models, although the detailed discussion of its specific effect on mass distribution should be the focus of another study. Meanwhile, we note that turbulence is not intrinsically in conflict with other mechanisms concerning dynamical encounters that contribute to specific features in $\chi_{\rm eff}$ distribution 
\citep{McKernan2022}, and they can be combined to give some robust and/or distinguishable prediction in the properties of GW events.}

\section*{Acknowledgements}
Y.X.C thanks Cl\'ement Baruteau for instructions on the simulation setup. We thank Hui Li, Adam Dempsey, Bhupendra Mishra and Yan-Fei Jiang for helpful discussions. We thank the anonymous referee for valuable suggestions that improved the clarity of the paper.

\section*{Data Availability}

The data underlying this paper will be shared on reasonable request to the corresponding author. 



\bibliographystyle{mnras}
\bibliography{mnras_template} 



\appendix

\section{Analytic approximation of initial spin decay}
\label{app:spin_decay}
To approximate the mean decay of spin magnitude, we consider an ideal ``alternating model" where the CSD spin flips direction every small fraction of the growth timescale $\Delta t  = \tau_M/\mathcal{S}$ with respect to BH spin, and $a$ switches deterministically between positive and negative values. These flips prevent the magnitude of the EBH spin from increasing monotonically towards $|a| \approx 1$, as in a basic accretion cycle. Instead, we can make $a$ fluctuate around zero without introducing RW diffusion factors. We define 

\begin{equation}
    \mathcal{R}_{\rm isco,\pm} = 3+Z_{2} \pm  \sqrt{\left(3-Z_{1}\right)\left(3+Z_{1}+2 Z_{2}\right)} \gtrless 6
\end{equation}
respectively on the spin-down (+) and spin-up (-) branches, so that we do not need to involve $\text{sign}()$ in taking derivatives. $Z_1, Z_2$ are defined in Equation \ref{Z1Z2}.

Consider two cycles starting from black hole mass $M_{\filledstar}$ and spin $a$, each cycle accreting a small amount of material approximately $\Delta M = M_{\filledstar} \Delta t/\tau_M$, on average a small amount will be chiseled off $|a|$ because spin down is a little more efficient than spin up. When $\mathcal{S}$ is large, 
the sign of vector $\mathbf{J_\filledstar}$ is not changed between two small cycles so the switch of $\mathbf{J_d}$ w.r.t. vertical direction is indistinguishable from that w.r.t. $\mathbf{J_\filledstar}$. As an example, take the first accretion cycle to be spinning down ($a<0$ regardless of $a_{abs}$), so we should adopt $\mathcal{R}_{\rm isco+}$ to 
calculate the reference ISCO radius:

\begin{equation}
\begin{aligned}
      \Delta a_1 & = M_{\filledstar} \dfrac{\Delta t}{{\tau_M}}\left.\dfrac{da}{dM_{\filledstar}}\right|^{M_0 = M_{\filledstar}}_{\mathcal{R}_{\rm isco, 0} = \mathcal{R}_{\rm isco+}(a)} ,
\end{aligned}
\label{eqn:da1}
\end{equation}
After EBH's spin w.r.t CSD grew to be $a + \Delta a_1$, but then the CSD flips so w.r.t CSD the EBH 
spin parameter becomes $-a - \Delta a_1$, and in this adjacent spin up phase the EBH gains 
\begin{equation}
\begin{aligned}
    \Delta a_2 & =  (M_{\filledstar}+\Delta M) \dfrac{\Delta t}{{\tau_M}}\left.\dfrac{da}{dM_{\filledstar}}\right|^{M_0 = M_{\filledstar}+\Delta M}_{\mathcal{R}_{\rm isco, 0 } = \mathcal{R}_{\rm isco-}(a + \Delta a_1)} .
\end{aligned}
\label{eqn:da2}
\end{equation}

After two cycles, the CSD spin switches again and $a$ becomes $a+\Delta a_1-\Delta a_2$. To first order, $\Delta a_1$ and $\Delta a_2$ scale linearly with $\Delta t$ since 

\begin{equation}
    \left.\frac{d a}{d M_{\filledstar}}\right|_{\mathcal{R}_{\mathrm{isco}, 0}=\mathcal{R}_{\mathrm{isco, \pm}}\left(a+\Delta a_{1}\right)}^{M_{0}=M_{\filledstar}+\Delta M} = \left.\frac{d a}{d M_{\filledstar}}\right|_{\mathcal{R}_{\mathrm{isco}, 0}=\mathcal{R}_{\mathrm{isco, \pm}}\left(a\right)} ^{M_{0}=M_{\filledstar}} + \mathcal{O}(\Delta t).
\end{equation}

Neglecting higher order terms, we have from Equation \ref{eqn:evolvea} that

\begin{equation}
\begin{aligned}
    \left.\dfrac{da}{dM_{\filledstar}}\right|_{\mathcal{R}_{\mathrm{isco}, 0}
    =\mathcal{R}_{\mathrm{isco, \pm}} \left(a\right)}^{M_{0}=M_{\filledstar}(t)}
    = \dfrac1{M_{\filledstar}}  (\dfrac{\mathcal{R}_{\rm isco,\pm}(a)^{3/2}}{(3 \mathcal{R}_{\rm isco,\pm}(a)-2)^{1/2}}- \\ \mathcal{R}_{\rm isco,\pm}(a)^{1/2}
    [4-(3 \mathcal{R}_{\rm isco,\pm}(a)-2)^{1/2}  ]) 
    := \dfrac{\mathcal{F}_{\pm}(a)}{M_\filledstar}.
    \end{aligned}
\end{equation}
The net change in $a$ during a spin-up followed by a spin-down event becomes
\begin{equation}
   \Delta a =  \Delta a_{1} - \Delta a_{2} = [\mathcal{F_+}(a) - \mathcal{F_-}(a)] \Delta t/\tau_M.
\end{equation}
For $a=0$, $[\mathcal{F_+}(a) - \mathcal{F_-}(a)]=0$, so that in the limit $|a| \ll 1$, 

\begin{equation}
    [\mathcal{F_+}(a) - \mathcal{F_-}(a)]=\left.\dfrac{d[\mathcal{F_+}(a) - \mathcal{F_-}(a)]}{da}\right|_{a=0} a + \mathcal{O}(a^2) 
    = - 5.752 a + \mathcal{O}(a^2).
    \label{eqn:smallaexpansion}
\end{equation}
which implies that the rate of change for $a$ on average is 
\begin{equation}
   \dfrac{\Delta a}{2\Delta t} \approx -2.876 a/\tau_M,
\end{equation}

When $\Delta t \rightarrow 0$ at $\mathcal{S}=\infty$, the asymptotic limit for $a$ evolution in the alternating case is to alternate between $\pm |a_0| e^{-2.876 t / \tau_{M}}$, while $|a|$ monotonically decreases. Although Equation \ref{eqn:smallaexpansion} is valid in the limit of small $|a|$, the exponential function can approximate the entire evolution very well since the initial decay of $|a|$ is quite rapid (See Figure \ref{fig:individual}, green dotted line). The decay time can be calculated as
$\tau_d \approx \tau_M\ln \mathcal{S}/5.752$, which is the time for $|a_0| = 1$ to reach $|a| ={\mathcal{S}}^{-0.5}$, while for other values of initial $|a_0|$ the decay time is even shorter. One may also introduce a shortening of $-0.14 \tau_M$ in $\tau_d$ for reaching the more accurate converged value $a_{RW} = 1.5{\mathcal{S}}^{-0.5}$ (Figure \ref{fig:pdf}) instead of  ${\mathcal{S}}^{-0.5}$, which is not significant. On a population level, $\tau_d$ marks the transition of an initial average spin-down phase towards the steady-state dominated by RW.

\section{Relevance to BBH Mergers and $\chi_{\rm eff}$ Distributions}
\label{app:chi_eff}

The detailed influence of an isotropic low-spin distribution on $\chi_{\rm eff}$ needs to be understood through population synthesis incorporating mass functions and more detailed physical effects. Nevertheless, we can 
offer some natural argument asserting that it's difficult for $\chi_{\rm eff}$ to reach values much larger 
than $a_{RW}$. Once captured into BBHs and after binary inspiral, each individual EBH with mass $M_1$ and $M_2$ 
makes a fractional contribution to $\chi_{\rm eff}$, such that

\begin{equation}
    \chi_{\rm eff} = \dfrac{M_1}{M_1+M_2}\chi_{1} + \dfrac{M_2}{M_1+M_2} \chi_{2}, \ \ \ \ \ \
    \chi_i = |a_i|\cos \psi_i, \ \ \ \ \ \ i=1,2
\end{equation}

where $\psi_i$ is angle between EBH spin and binary orbital axis. Avoiding making any 
specific assumptions about the mass distribution, we understand that in the limit that the evolution towards merger is short, 
we can draw the magnitude of $|a|$ from the positive part of a Gaussian distribution $f(|a|)$, and $\lambda = \cos\psi $ from a uniform distribution $g(\lambda)$ since the spins have yet to couple with the orbital angular momentum. Neglecting all normalization coefficients, we have the cumulative probability distribution (CDF) for positive $\chi= |a|\lambda$ being

\begin{equation}
\begin{aligned}
    CDF(0< \chi < |a|\lambda) \propto \int_\chi^1 f(|a|)da\int_{\chi/|a|}^1 g(\lambda)d\lambda \\ \propto \int_\chi^1 f(|a|) \left(1-\dfrac{\chi}{|a|} \right)da
    \end{aligned}
\end{equation}
Substituting the exact form of $f(|a|)$, we can derive the probability distribution (PDF) of $\chi$ by differentiating the CDF

\begin{equation}
\begin{aligned}
PDF(\chi) \propto
-\dfrac{d}{d\chi}\int_\chi^{1} f(|a|)\left(1-\dfrac{\chi}{|a|}\right)d|a| = \dfrac{d}{d\chi}\int_\chi^{1} f(|a|)\left(\dfrac{\chi}{|a|}\right)d|a| \\ \approx \int_\chi^{\infty}
\exp\left({\dfrac{-|a|^2}{2 a_{RW}^2}}\right)\dfrac{d|a|}{|a|} \propto \mathcal{E}\left(\dfrac{\chi^2}{2 a_{RW}^2}\right),
\end{aligned}
\end{equation}

\begin{equation}
    \mathcal{E}(z) := \int_1^\infty \dfrac{e^{-xz} dx}{x} =\int_z^\infty  \dfrac{e^{-x}}{x}dx.
\end{equation}

Note we have approximated the upper limit to be $\infty$ since the integral from 1 to infinity is also negligible. Since the $\chi<0$ distribution is completely symmetric, after normalization we have:

\begin{equation}
    PDF(\chi) \approx \dfrac1{2\sqrt{2\pi} a_{RW}} \mathcal{E}(\dfrac{\chi^2}{2 a_{RW}^2}).
\end{equation}
This distribution function is very narrow and suggests that $\chi$ has $\sim 90.6\%$ probability of being between $\pm a_{\rm RW}$. Since for any general EBH mass function, the mass-weighted average $\chi_{\rm eff}$ cannot exceed the value of $\chi$ by much, we can constrain the magnitude of $\chi_{\rm eff}$ produced from the turbulence channel to be $\lesssim a_{RW}$ for the decoupled spin-orbit angular momentum scenario.

\bsp	
\label{lastpage}
\end{document}